\documentclass[a4paper,fleqn,usenatbib,useAMS]{mnras}
\usepackage[T1]{fontenc}
\usepackage{ae,aecompl}
\usepackage{graphicx}
\usepackage{amsmath}
\usepackage{amssymb}
\usepackage{lineno}

\newcommand\xmm{{\it XMM-Newton}}

\newcommand\swift{{\it Swift}}
\newcommand\nustar{{\it NuSTAR}}
\newcommand\s{{\rm~s}}
\newcommand\ks{{\rm~ks}}

\newcommand\hz{{\rm~Hz}}
\newcommand\keV{{\rm~keV}}
\newcommand\ev{{\rm~eV}}

\title[High-density disc reflection in Mrk~1044]{A high-density relativistic reflection origin for the soft and hard X-ray excess emission from Mrk~1044}

\author[L. Mallick et al.]{L. Mallick$^{1}$,\thanks{Email: labani@iucaa.in (LM)}
	W. N. Alston$^{2}$,\thanks{Email: wna@ast.cam.ac.uk (WNA)}
	M. L. Parker$^{3}$,\thanks{Email: mparker@sciops.esa.int (MLP)}
	A. C. Fabian$^{2}$,	
 C. Pinto$^{2}$, 
\newauthor G. C. Dewangan$^{1}$,
A. Markowitz$^{4,5}$,
P. Gandhi$^{6}$,            	          	
	A. K. Kembhavi$^{1}$,
	 R. Misra$^{1}$ \\
$^{1}$ Inter-University Centre for Astronomy and Astrophysics, Post Bag 4, Ganeshkhind, Pune 411007, India\\
$^{2}$ Institute of Astronomy, University of Cambridge, Madingley Road, Cambridge CB3 0HA, UK\\
$^{3}$ European Space Astronomy Centre (ESA/ESAC), E-28691 Villanueva de la Ca$\tilde{n}$ada, Madrid, Spain\\
$^{4}$ Nicholas Copernicus Astronomical Center, ul. Bartycka 18, PL-00-716 Warsaw, Poland\\
$^{5}$ Center for Astrophysics and Space Sciences, Univ. of California, San Diego, MC 0424, La Jolla, California, 92093-0424, U.S.A. \\
$^{6}$ School of Physics and Astronomy, University of Southampton, Highfield, Southampton SO17 1BJ, UK 
}

\begin{document}
\date{\today}

\pagerange{\pageref{firstpage}--\pageref{lastpage}} 

\maketitle


\label{firstpage}

\begin{abstract}
We present the first results from a detailed spectral-timing analysis of a long ($\sim$130\ks{}) \xmm{} observation and quasi-simultaneous \nustar{} and \swift{} observations of the highly-accreting narrow-line Seyfert~1 galaxy Mrk~1044. The broadband (0.3$-$50\keV{}) spectrum reveals the presence of a strong soft X-ray excess emission below $\sim$1.5\keV{}, iron~K$_{\alpha}$ emission complex at $\sim$6$-$7\keV{} and a `Compton hump' at $\sim$15$-$30\keV{}. We find that the relativistic reflection from a high-density accretion disc with a broken power-law emissivity profile can simultaneously explain the soft X-ray excess, highly ionized broad iron line and the Compton hump. At low frequencies ($[2-6]\times10^{-5}$\hz{}), the power-law continuum dominated 1.5$-$5\keV{} band lags behind the reflection dominated 0.3$-$1\keV{} band, which is explained with a combination of propagation fluctuation and Comptonization processes, while at higher frequencies ($[1-2]\times10^{-4}$\hz{}), we detect a soft lag which is interpreted as a signature of X-ray reverberation from the accretion disc. The fractional root-mean-squared (rms) variability of the source decreases with energy and is well described by two variable components: a less variable relativistic disc reflection and a more variable direct coronal emission. Our combined spectral-timing analyses suggest that the observed broadband X-ray variability of Mrk~1044 is mainly driven by variations in the location or geometry of the optically thin, hot corona.
\end{abstract}


\section{Introduction}
The basic mechanism underlying the activity of active galactic nuclei (AGN) is the accretion of matter onto the central supermassive black holes (SMBHs) of mass $M_{\rm BH}\sim10^{5}-10^{10} M_{\rm \odot}$. AGN are the most luminous ($10^{41}-10^{47}$erg~s$^{-1}$) persistent sources of electromagnetic radiation in the universe and have been the centre of interest for several reasons. They are considered as the only probes to determine the changing demographics and accretion history of SMBHs with cosmic time. The evolution of galaxies is closely linked with the growth and energy output from SMBHs (e.g. \citealt{kh13}). The feedback from AGN can regulate the star formation in their host galaxies and hence play a key role in galaxy evolution (see e.g. \citealt{fab12}). 

AGN emit over the entire electromagnetic spectrum, from radio to gamma-rays, with each waveband allowing us to probe different aspects of the physics of accretion onto the central SMBH. The broadband X-ray spectrum of a large fraction of AGN generally consists of the following main components: a primary power-law continuum, a soft X-ray excess emission below $\sim2$\keV{}, iron (Fe)~K emission complex at around 6$-$7\keV{}, Compton hump above $10$\keV{} and complex absorption. The power-law continuum is thought to be produced by inverse-Compton scattering of lower energy optical/UV seed photons from the accretion disc in a corona consisting of hot electrons \citep{st80,ha91,ha93}. The X-ray continuum emission from the hot corona is bent down onto the accretion disc due to strong gravity and gives rise to X-ray reflection features by the processes of Compton scattering, photoelectric absorption, fluorescent line emission and bremsstrahlung \citep{rf05}. The most notable reflection feature is the Fe~K complex in the 6$-$7\keV{} energy range. The reflection spectrum is blurred by the combination of Doppler shifts, relativistic beaming and gravitational redshifts in the strong gravitational field very close to the black hole \citep{fa89,la91}. Moreover, we observe absorption features in the X-ray spectrum which can be caused by the presence of absorbing clouds along the line-of-sight or wind launched from the surface of the accretion disc (e.g. \citealt{po03,to11,re14,par17,pin18}).

The broad Fe~K$_{\alpha}$ emission line and soft X-ray excess emission are considered as direct probes available for the innermost accretion disc and black hole spin. However, the physical origin of the soft excess (SE) emission in many AGN remains highly debated (e.g. \citealt{cr06,do12}). The SE can be modelled physically by thermal Comptonization of the optical/UV seed photons in an optically thick, warm corona (e.g. \citealt{de07,lo12,po18}) and/or relativistically broadened lines from the inner ionized accretion disc (e.g. \citealt{cr06,fa09,na11,ga14}). Recently, \citet{ga16} have attempted to solve this issue by providing a new reflection model where the density of the disc atmosphere is a free parameter varying over the range $n_{\rm e}=10^{15}-10^{19}$~cm$^{\rm -3}$. The main effect of high-density on the reflection model occurs below 3\keV{} and is very important for powerful coronae and low-mass, highly-accreting black holes. The high-density relativistic reflection model has been successfully applied in one black-hole binary Cygnus~X-1 \citep{tom18} and one AGN IRAS~13224--3809 \citep{ji18}.

AGN are variable in all wavebands over timescales from a few seconds to years depending upon the physical processes they are governed by. The X-ray emission from AGN is ubiquitous and shows strong variability which depends on energy and flux of the source (e.g. \citealt{mark07,va11,al13,par15,ma17,al18}). This strong variability implies that the X-ray emission is originated in the inner regions of the central engine. Despite many efforts to understand the variable AGN emission, the exact origin of the energy-dependent variability of AGN is not clearly understood because of the presence of multiple emission components and the interplay between them. There are a few approaches that can probe the nature and origin of energy-dependent variability in AGN. One compelling approach is to measure the time-lag between different energy bands of X-ray emission (e.g. \citealt{pa01,mc04,ar06,ams08}). The measured time lag could be both frequency and energy dependent (e.g. \citealt{dm11,ka14}). Depending on the source geometry and emission mechanism, the time lags have positive and/or negative values. The positive or hard lag refers to the delayed hard photon variations relative to the soft photons and is detected at relatively low frequencies. The hard lag was first seen in X-ray binaries (e.g. \citealt{mi88,no99,ko01}). The origin of positive or hard lag is explained in the framework of viscous propagation fluctuation scenario \citep{ko01,au06,ut11,hogg16}. The negative or soft time-lag (soft photon variations are delayed with respect to the hard photons) is usually observed at higher frequencies. The soft lag was first discovered in one low-mass bright AGN 1H~0707--495 \citep{fa09} and interpreted as a sign of the reverberation close to the black hole (e.g. \citealt{zo10,em11,zo12,ca13,ka14}). The alternative justification for soft lag is the reflection from distant clouds distributed along the line sight or the accretion disc wind \citep{mi10}.

Another important perspective to probe the origin of energy-dependent variability is to examine the fractional root-mean-squared (rms) variability amplitude as a function of energy, the so-called fractional rms spectrum \citep{ed02,va03}. The modelling of the fractional rms spectra can shed light on the variable emission mechanisms and has been performed in a handful of AGN (e.g. MCG--6-30-15: \citealt{mi07}, 1H~0707--495: \citealt{fa12}, RX~J1633.3+4719: \citealt{ma16}, PG~1404+226: \citealt{md17}, Ark~120: \citealt{ma17}). This tool acts as a bridge between the energy spectrum and the observed variability and can effectively identify the constant and variable emission components present in the observed energy spectrum. It provides an orthogonal way to probe the variable components and the causal connection between them. Moreover, the Fourier frequency-resolved fractional rms spectrum allows us to understand both frequency and energy dependence of variability and hence different physical processes occurring on various timescales.

Here we perform the broadband (0.3$-$50\keV{}) spectral and timing studies of the highly-accreting AGN Mrk~1044 using data from \xmm{}, \swift{} and \nustar{}. We explore the new high-density relativistic reflection model as the origin of both soft and hard X-ray excess emission as well as the underlying variability mechanisms in the source. Mrk~1044 is a radio-quiet narrow-line Seyfert~1 galaxy at redshift $z=0.016$. The central SMBH mass of Mrk~1044 obtained from the H$_\beta$ reverberation mapping is $3\times10^6 M_{\rm \odot}$ \citep{wa01,du15}. The dimensionless mass accretion rate as estimated by \citet{du15} using the standard thin disc equation is $\dot{m}=\frac{\dot{M}c^{2}}{L_{\rm E}}=20.1\left(\frac{l_{44}}{cosi}\right)^{3/2}\left(\frac{M_{BH}}{10^{7}M_{\odot}}\right)^{-2}=16.6^{+25.1}_{-10.1}$, where $L_{\rm E}$ is the Eddington luminosity, $M_{BH}$ is the SMBH mass, $i$ is the disc inclination angle, $l_{44}=L_{5100}/10^{44}$~erg~s$^{-1}$ where $L_{5100}$ represents the AGN continuum luminosity at the rest frame wavelength of $5100\textrm{\AA}$.

The paper is organized as follows. In Section~\ref{sec:obs}, we describe the observations and data reduction method. In Section~\ref{sec:spec}, we present the broadband (0.3$-$50\keV{}) spectral analysis and results. In Section~\ref{sec:time}, we present the timing analysis and results. In Section~\ref{sec:EDV}, we present the energy-dependent variability study of the source in different Fourier frequencies. We summarize and discuss our results in Section~\ref{sec:discussion}.

\section{Observations and Data Reduction}
\label{sec:obs}
\subsection{\xmm{}}
\xmm{} \citep{ja01} observed Mrk~1044 on 27th January 2013 (Obs.~ID 0695290101) with a total duration of $\sim130$\ks{}. We analyzed the data from the European Photon Imaging Camera (EPIC-pn; \citealt{st01}) and Reflection Grating Spectrometer (RGS; \citealt{den01}). The EPIC-pn camera was operated in the small window (\textsc{sw}) mode with the thin filter to decrease pile up. The log of \xmm{}/EPIC-pn observation used in this work is shown in Table~\ref{table0}. The data sets were processed with the Scientific Analysis System (\textsc{sas}~v.15.0.0) and the updated (as of 2017 December 31) calibration files. We processed the EPIC-pn data using \textsc{epproc} and produced calibrated photon event files. We filtered the processed pn events with \textsc{pattern}$\leq4$ and \textsc{flag}$==0$ by taking both single and double pixel events but removing bad pixel events. To exclude the proton flare intervals, we created a \textsc{gti} (Good Time Interval) file above 10\keV{} for the full field with the \textsc{rate}$<0.5$\rm~cts~s$^{-1}$ using the task \textsc{tabgtigen} and acquired the maximum signal-to-noise. We examined for the pile-up with the task \textsc{epatplot} and did not find any pile-up in the EPIC-pn data. We extracted the source spectrum from a circular region of radius 30~arcsec centred on the source and the background spectrum from a nearby source free circular region with a radius of 30~arcsec. We produced the Redistribution Matrix File (\textsc{rmf}) and Ancillary Region File (\textsc{arf}) with the \textsc{sas} tasks \textsc{rmfgen} and \textsc{arfgen}, respectively. We extracted the background-subtracted, deadtime and vignetting corrected source light curves for different energy bands from the cleaned pn event file using the task \textsc{epiclccorr}. Finally, we grouped the EPIC-pn spectral data in order to oversample by at least a factor of 5 and to have a minimum of 20 counts per energy bin with the task \textsc{specgroup}. The net count rate estimated for EPIC-pn is ($17.48\pm0.01$)\rm~cts~s$^{-1}$ resulting in a total of $1.26\times10^{6}$ pn counts. 

We processed the RGS data with the task \textsc{rgsproc}. The response files were produced using the \textsc{sas} task \textsc{rgsrmfgen}. We combined the spectra and response files for RGS~1 and RGS~2 using the task \textsc{rgscombine}. Finally, we grouped the RGS spectral data with a minimum of 25 counts per energy bin using the \textsc{ftools} \citep{bl95} task \textsc{grppha}.

\subsection{\nustar{}}
Mrk~1044 was observed with the \nustar{} telescope \citep{ha13} on February 8, 2016 (Obs.ID~60160109002) starting at 07:31:08 UT with a total duration of $\sim40$\ks{} with its two co-aligned focal plane modules FPMA and FPMB. The observation log of \nustar{} is shown in Table~\ref{table0}. We analyzed the data sets with the \nustar{} Data Analysis Software (\textsc{nustardas}) package (v.1.6.0) and the updated (as of 2017 December 31) calibration data base (version 20170222). We produced the calibrated and cleaned photon event files with the task \textsc{nupipeline}. We used the script \textsc{nuproducts} to extract the source spectra and light curves from a circular region of radius 50~arcsec centred on the source while the background spectrum and light curves were extracted from two same-sized circular regions free from source contamination and the edges of the CCD. We generated the background-subtracted FPMA and FPMB light curves with the \textsc{ftools} task \textsc{lcmath}. Finally, we grouped the spectra using the \textsc{grppha} tool with a minimum of 30 counts per bin. The net count rate is $\sim0.18$\rm~cts~s$^{-1}$ resulting in a total of $\sim4000$ counts with the net exposure time of about 22\ks{} for both FPMA and FPMB. 

\subsection{\swift{}}
\swift{} \citep{ge04} observed Mrk~1044 several times with one observation quasi-simultaneous with \nustar{} on February 8, 2016 (Obs.ID~00080912001) starting at 12:09:57 UT for a total duration of $\sim6.2$\ks{}. The details of the \swift{} observations used in this work are listed in Table~\ref{table0}. Here we analyzed the data from the X-Ray Telescope (XRT; \citealt{bu05}) with the XRT Data Analysis Software (\textsc{xrtdas}) package (v.3.2.0) and the recent (as of 2017 December 31) calibration data base (version 20171113). We generated the cleaned event files with the task \textsc{xrtpipeline}. We have extracted the source spectra from a circular region of radius 40~arcsec centred on the source position and background spectra were produced from a nearby source free circular region of radius 40~arcsec with the script \textsc{xrtproducts}. We combined the spectra and response files for all observations with the tool \textsc{addascaspec}. Finally, we grouped the spectral data with the \textsc{grppha} tool so that we had a minimum of 20 counts per energy bin. The total net XRT count rate is $\sim0.6$\rm~cts~s$^{-1}$ resulting in a total of $\sim16500$ counts with the net exposure time of $\sim29.4$\ks{}. 
 
 \begin{table*}
  \centering
\caption{\xmm{}/EPIC-pn, \nustar{} and \swift{}/XRT observations of Mrk~1044. The net exposure time is the live time of the instrument after background filtering.}
\begin{center}
\scalebox{1.0}{%
\begin{tabular}{cccccccc}
\hline 
Satellite & Camera & Obs. ID & Date & Net exposure & Net count rate \\
            & & & (yyyy-mm-dd)  & (ks) & (counts~s$^{-1}$) \\                                      
\hline 
\xmm{} & EPIC-pn & 0695290101 & 2013-01-27 & 72.0 & 17.5 \\ [0.2cm]
\hline 
\nustar{} & FPMA  & 60160109002 & 2016-02-08 & 21.7 & 0.18 \\ [0.2cm]
          & FPMB  & 60160109002 & 2016-02-08 & 21.6 & 0.18 \\ [0.2cm]
\hline 
\swift{} & XRT  & 00035760001 & 2007-07-25  & 3.0 & 0.71 \\ [0.2cm]
         &      & 00035760002 & 2007-08-01  & 7.6 & 0.32 \\ [0.2cm]
         &      & 00035760003 & 2008-03-02  & 5.5 & 0.21 \\ [0.2cm]
         &      & 00091316001 & 2012-06-05  & 3.0 & 1.66 \\ [0.2cm]
         &      & 00080912001 & 2016-02-08  & 6.1 & 0.59 \\ [0.2cm]
         &      & 00093018003 & 2017-09-08  & 4.1 & 0.72 \\ [0.2cm]
\hline 
\end{tabular}}
\end{center} 
\label{table0}           
\end{table*}

\begin{figure*}
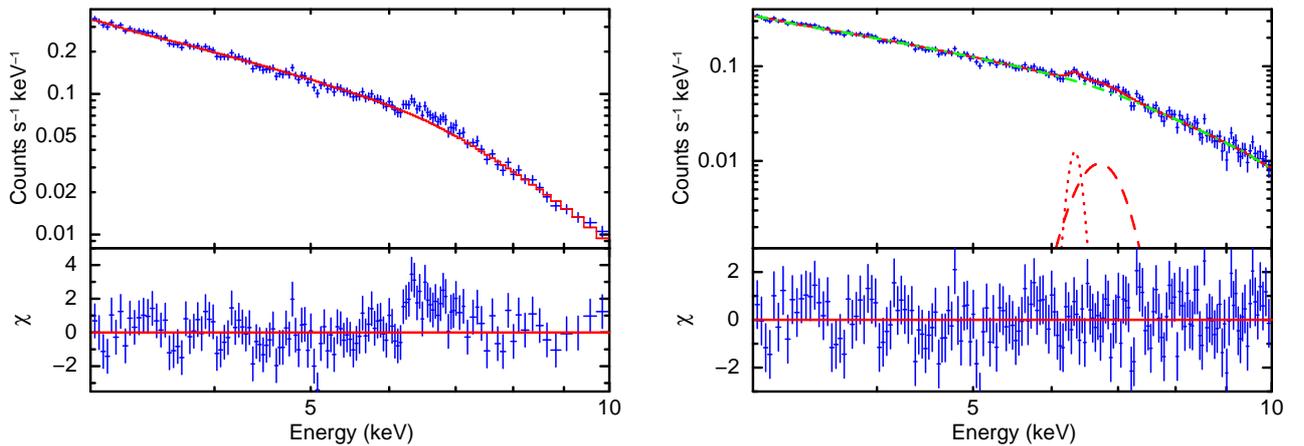

\includegraphics[scale=0.32,angle=-90]{fig1a.ps}
\includegraphics[scale=0.32,angle=-90]{fig1b.ps}
\caption{Left: The \xmm{}/EPIC-pn 3$-$10\keV{} spectrum, the \textsc{zpowerlw} model ($\Gamma = 1.96$) modified by the Galactic absorption and the residuals. A strong residual in the Fe~K region (6$-$7\keV{}) is clearly seen. The data are binned up for clarity. Right: The EPIC-pn 3$-$10\keV{} spectrum, the best-fitting phenomenological model, \textsc{tbabs$\times$(zgauss1$+$zgauss2$+$zpowerlw)} and the residual spectrum. The model consists of an absorbed power-law along with narrow and broad iron emission lines.}
\label{pn_hard_pow}
\end{figure*}

\begin{figure*}
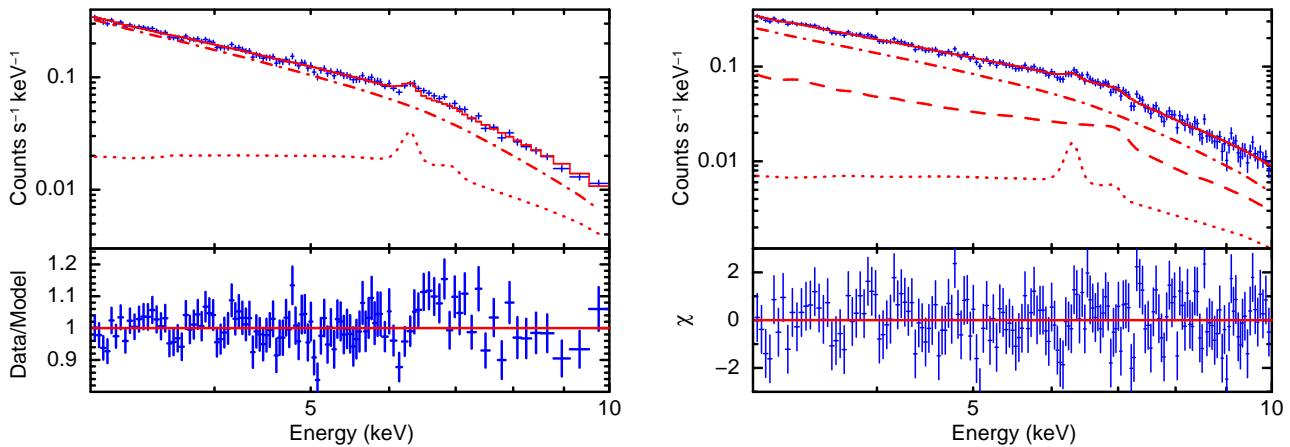

\includegraphics[scale=0.32,angle=-90]{fig2a.ps}
\includegraphics[scale=0.32,angle=-90]{fig2b.ps}
\caption{Left: The EPIC-pn 3$-$10\keV{} spectral data, the fitted distant reflection model [\textsc{tbabs$\times$(xillverd+zpowerlw)}] and the residuals which show a broad excess emission in the $6.5-6.9$\keV{} energy range. The data are binned up for clarity. Right: The EPIC-pn 3$-$10\keV{} spectrum, the best-fitting reflection model, \textsc{tbabs$\times$(relxilld+xillverd+zpowerlw)} along with the model components and the residuals. The best-fitting model consists of three main components: a primary power-law emission (in dash-dotted), an ionized, relativistic disc reflection with a single power-law emissivity profile (in dashed line) and a neutral, distant reflection (in dotted line).}
\label{pn_hard_best}
\end{figure*}

\begin{figure*}
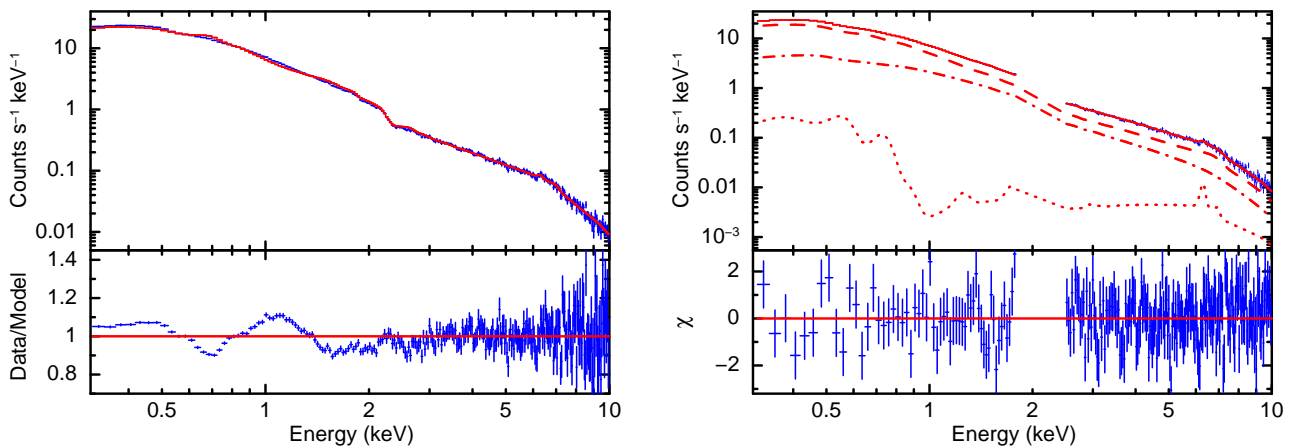

\includegraphics[scale=0.32,angle=-90]{fig3a.ps}
\includegraphics[scale=0.32,angle=-90]{fig3b.ps}
\caption{Left: The full band (0.3$-$10\keV{}) EPIC-pn spectrum, the hard band (3$-$10\keV{}) best-fitting spectral model, \textsc{tbabs$\times$(relxilld+xillverd+zpowerlw)} extrapolated down to 0.3\keV{} and the ratio of data to the model. Right: The full band (0.3$-$10\keV{}) EPIC-pn spectrum, the best-fitting model, \textsc{tbabs$\times$gabs$\times$(relxilld+xillverd+zpowerlw)} and the residuals. The model components are: a primary power-law emission (in dash-dotted), an ionized, high-density relativistic disc reflection with a broken power-law emissivity profile (in dashed), a neutral distant reflection (in dotted), and a broad Gaussian absorption line at around 0.7\keV{}.}
\label{pn_ldr}
\end{figure*}

\section{Broadband (0.3$-$50\keV{}) Spectral analysis: A first look}
\label{sec:spec}
We performed the spectral analysis of Mrk~1044 in \textsc{xspec}~v.12.9.0n \citep{ar96} and \textsc{spex}\footnote[1]{\url{https://www.sron.nl/spex}}~v.3.04.00 \citep{ka96}. We employed the $\chi^{2}$ statistics and estimated the errors at the 90~per~cent confidence limit corresponding to $\Delta\chi^{2}=2.71$, unless specified otherwise.  

\subsection{The 3$-$10\keV{} EPIC-pn spectrum}
AGN are well known to exhibit a complex soft X-ray excess below $\sim2$\keV{} and hence we first focused on the hard X-ray emission above 3\keV{}. We begin the spectral analysis by fitting the 3$-$10\keV{} EPIC-pn spectrum of Mrk~1044 with a simple power-law (\textsc{zpowerlw}) corrected by the Galactic absorption model (\textsc{tbabs}) assuming the cross-sections and solar interstellar medium (ISM) abundances of \citet{aspl09}. The Galactic neutral hydrogen column density was fixed at $N_{\rm H}=3.6\times10^{20}$\rm~cm$^{-2}$ \citep{wil13}. The fitting of the 3$-$10\keV{} data with the \textsc{tbabs$\times$zpowerlw} model provided an unacceptable fit with $\chi^{2}$/d.o.f = 220.2/165 and a strong residual in the Fe~K region (6$-$7\keV{}) as shown in Figure~\ref{pn_hard_pow} (left). In order to assess the presence of a neutral Fe~K emission from the `torus' or other distant material, we model an emission line that is expected to be unresolved by the EPIC-pn camera. Any additional broad contribution to the line profile would then come from material closer to the black hole, e.g., the inner disc. Therefore, we first added one narrow Gaussian line (\textsc{zgauss1}) to model the neutral Fe K emission by fixing the line width at $\sigma_{N}=10$\ev{}, which improved the fit statistics to $\chi^{2}$/d.o.f = 186.6/163 ($\Delta\chi^{2}$=$-$33.6 for 2 d.o.f). However, the residual plot shows a broad excess emission centered at $\sim6.8$\keV{}. Then, we added another Gaussian line (\textsc{zgauss2}) and set the line width to vary freely. The fitting of the 3$-$10\keV{} EPIC-pn spectrum with the phenomenological model, \textsc{tbabs$\times$(zgauss1$+$zgauss2$+$zpowerlw)} provided a statistically acceptable fit with $\chi^{2}$/d.o.f = 154.3/160 ($\Delta\chi^{2}$=$-$32.3 for 3 d.o.f) as shown in Fig.~\ref{pn_hard_pow} (right). The centroid energies of the narrow and broad emission lines are $E_{1}=6.43^{+0.05}_{-0.04}$\keV{} and $E_{2}=6.84^{+0.14}_{-0.15}$\keV{}, which are representatives of the neutral Fe~K$_{\alpha}$ line and highly-ionized Fe line, respectively. The line width of the ionized Fe emission line is $\sigma_{B}=0.31^{+0.23}_{-0.14}$\keV{}. The best-fitting values for the equivalent width of the Fe~K$_{\alpha}$ and ionized Fe emission lines are ${\rm EW_{1}}=36.9^{+24.8}_{-23.5}$\ev{} and ${\rm EW_{2}}=131.7^{+57.2}_{-62.8}$\ev{}, respectively. Thus, we infer that the 3$-$10\keV{} EPIC-pn spectrum of Mrk~1044 is well described by a power-law like primary emission along with neutral, narrow and ionized, broad iron emission lines. 

Since modelling of the iron emission features with Gaussian lines is not physically realistic, we explored the physically-motivated reflection models \citep{ga14,ga16} to fit the iron emission lines and hence understand the physical conditions of the accretion disc. First, we modelled the narrow Fe~K$_{\alpha}$ emission feature with the reflection (\textsc{xillverd}) model \citep{ga16} without relativistic blurring, as appropriate for distant reflection and with density fixed at $n_{\rm e}=10^{15}$cm$^{\rm -3}$, the lowest value afforded by the model. We also fixed the ionization parameter of the \textsc{xillverd} model at its minimum value ($\log\xi=0$) as the Fe~K$_{\alpha}$ line is neutral. The inclination angle of the distant reflector was fixed at a higher value, $i=60^{\circ}$ and decoupled from the relativistically blurred reflection component modelling the disc reflection. Since the incident continuum as assumed by the high-density reflection model is a power-law {\it without} a variable high-energy cut-off parameter, we have used a simple power-law (\textsc{zpowerlw}) model as the illuminating continuum. The fitting of the 3$-$10\keV{} spectrum with the model, \textsc{tbabs$\times$(xillverd+zpowerlw)} provided a $\chi^{2}$/d.o.f = 184.1/163 with a broad excess emission at $\sim6.5-6.9$\keV{} energy range (see Figure~\ref{pn_hard_best}, left). The origin of the broad Fe line could be the relativistic reflection from the inner regions of an ionized accretion disc. Therefore, we have added the high-density relativistic reflection model (\textsc{relxilld}; \citealt{ga16}) with a single power-law emissivity profile. The relevant parameters of the \textsc{relxilld} model are: emissivity index ($q$, where emissivity of the relativistic reflection is defined by $\epsilon\propto r^{-q}$), inner disc radius ($r_{\rm in}$), outer disc radius ($r_{\rm out}$), black hole spin ($a$), disc inclination angle ($\theta^{\circ}$), ionization parameter, electron density, iron abundance, reflection fraction ($R$) and photon index. We fixed the SMBH spin and outer disc radius at $a=0.998$ and $r_{\rm out}=1000r_{\rm g}$, respectively, where $r_{\rm g}=GM_{\rm BH}/c^{2}$. The reflection fraction of the \textsc{relxilld} model was fixed at $R=-1$ to obtain only the reflected emission. The photon index and iron abundance for the relativistic reflection (\textsc{relxilld}) component were tied with the distant reflection (\textsc{xillverd}) component. In \textsc{xspec}, the 3$-$10\keV{} model reads as \textsc{tbabs$\times$(relxilld+xillverd+zpowerlw)} which provided a reasonably good fit with $\chi^{2}$/d.o.f = 157.2/157. The best-fitting values for the photon index, emissivity index, disc inclination angle, ionization parameter and electron density are $\Gamma=2.25^{+0.07}_{-0.09}$, $q=10^{+0p}_{-5.7}$, $\theta^{\circ}=44.5^{+5.2}_{-2.3}$, $\xi= 492^{+223}_{-382}$~erg~cm~s$^{-1}$ and $\log(n_{\rm e}$/cm$^{\rm -3})=16.4^{+2.6p}_{-1.4p}$, respectively. Therefore, we infer that the hard X-ray (3$-$10\keV{}) spectrum of Mrk~1044 is well explained by a neutral distant reflection as well as an ionized, relativistic disc reflection where the emission is centrally concentrated in the inner regions of the accretion disc. The 3$-$10\keV{} EPIC-pn spectral data, the best-fitting physical model along with all the model components and the deviations of the observed data from the model are shown in Fig.~\ref{pn_hard_best} (right).  

\begin{table*}
 \centering
 \caption{The best-fitting model parameters for the EPIC-pn (0.3$-$10\keV{}) spectrum. Parameters with notations `(f)' and `$\ast$' indicate fixed and tied values, respectively. Errors are quoted at a 90~per~cent confidence level and estimated from the MCMC output.}
\begin{center}
\scalebox{0.95}{%
\begin{tabular}{cccccc}
\hline
Component & Parameter & EPIC-pn  & Description \\
 & & (0.3$-$10\keV{}) & \\ [0.2cm]
\hline 
Galactic absorption (\textsc{tbabs}) & $N_{\rm H}$(10$^{20}$~cm$^{-2}$) &  $4.6^{+0.2}_{-0.3}$ & Galactic neutral hydrogen column density \\[0.2cm]
Intrinsic absorption (\textsc{gabs}) & $E_{\rm abs}$(eV) & $705.5^{+8.3}_{-12.6}$ & Absorption line energy in the observed ($z=0.016$) frame \\ [0.2cm]
             & $\sigma_{\rm abs}$(eV)  & $77.5^{+15.8}_{-8.0}$  & Absorption line width \\ [0.2cm]             
       & $\tau_{\rm abs}$(10$^{-2}$)  & $1.8^{+0.6}_{-0.2}$ & Absorption line depth  \\ [0.2cm]
Relativistic reflection (\textsc{relxilld}) & $q_{\rm in}$ &  $9.73^{+0.24}_{-1.13}$ & Inner emissivity index\\ [0.2cm] 
                  & $q_{\rm out}$ &  $2.4^{+0.5}_{-0.2}$ & Outer emissivity index\\ [0.2cm]
                  & $r_{\rm br}$($r_{\rm g}$) &  $3.2^{+0.3}_{-0.3}$ & Break disc radius\\ [0.2cm]
                 & $a$ &  $0.992^{+0.005}_{-0.016}$ & SMBH spin  \\ [0.2cm]  
                 & $\theta^{\circ}$  & $46.4^{+1.9}_{-5.0}$ & Disc inclination angle\\ [0.2cm]
 & $r_{\rm in}$($r_{\rm g}$) &  $1.29^{+0.12}_{-0.05}$ & Inner disc radius\\ [0.2 cm]
  & $r_{\rm out}$($r_{\rm g}$) & 1000(f) & Outer disc radius \\ [0.2cm] 
 & $\Gamma$  & 2.29$^{\ast}$& Blurred reflection photon index  \\ [0.2cm]       
                 & $A_{\rm Fe}$  & 2.2$^{+0.5}_{-0.6}$& Iron abundance (solar)\\ [0.2cm]    
                 & $\log(n_{\rm e}$/cm$^{\rm -3}$)  & 16.2$^{+0.3}_{-0.1}$& Electron density of the disc \\ [0.2cm]    
                 & $\xi_{\rm blur}$(erg~cm~s$^{-1}$)  & 909$^{+90}_{-201}$&  Ionization parameter for the disc\\ [0.2cm]       
           & $N_{\rm blur}$(10$^{-4}$) &  2.4$^{+0.1}_{-1.1}$ &  Normalization of the relativistic reflection component \\ [0.2cm]        
Distant reflection (\textsc{xillverd}) & $\Gamma$ & 2.29$^{\ast}$ & Distant reflection photon index\\ [0.2cm]       
                 & $A_{\rm Fe}$ & 2.2$^{\ast}$ & Iron abundance (solar)\\ [0.2cm]    
                 & $\log(n_{\rm e}$/cm$^{\rm -3}$) & 15(f) & Density of the distant reflector \\ [0.2cm]    
                & $\xi_{\rm distant}$(erg~cm~s$^{-1}$)& 1(f) & Ionization parameter of the distant reflector\\ [0.2cm]       
                & $i^{\circ}$  & 60(f) & Inclination angle of the distant reflector  \\ [0.2cm]     
               & $N_{\rm distant}$(10$^{-6}$) & 4.6$^{+2.4}_{-2.3}$& Normalization of the distant reflection component  \\ [0.2cm]   
Incident continuum (\textsc{zpowerlw}) & $\Gamma$ & 2.29$^{+0.01}_{-0.03}$ & Photon index of the incident continuum \\ [0.2cm]
       & $N_{\rm PL}$(10$^{-3}$) & 2.1$^{+0.9}_{-0.5}$ & Normalization of the incident continuum \\ [0.2cm]  
\textsc{flux} & $F_{0.3-2}$(10$^{-11}$) & 2.26  & Observed 0.3$-$2\keV{} flux in units of erg~cm$^{-2}$~s$^{-1}$ \\ [0.2cm]   
     & $F_{2-10}$(10$^{-11}$) & 1.07 & Observed 2$-$10\keV{} flux in units of erg~cm$^{-2}$~s$^{-1}$ \\ [0.2cm]  
     & $\chi^2$/$\nu$ & 239.5/225 & Fit statistic  \\ [0.2cm]  
\hline     
\end{tabular}}
\end{center}
\label{table1}
\end{table*}

\begin{table*}
 \centering
 \caption{The best-fitting physical model parameters for the Galactic and intrinsic absorptions, which are obtained from the fitting of the RGS (0.38$-$1.8\keV{}) spectrum.}
\begin{center}
\scalebox{0.95}{%
\begin{tabular}{cccccc}
\hline
Component&Parameter & RGS (0.38$-$1.8\keV{}) &Description \\
\hline 
Galactic absorption (\textsc{hot}) & $N_{\rm H}$(10$^{20}$~cm$^{-2}$) &  $3.8^{+0.1}_{-0.1}$ & Galactic neutral hydrogen column density \\[0.2cm]
Intrinsic absorption (\textsc{xabs}) & $v_{\rm out}$($c$)  & $0.1^{+0.01}_{-0.01}$ & Outflow velocity of the wind  \\ [0.2cm]     
              & $N_{\rm H}^{\rm abs}$(10$^{20}$~cm$^{-2}$) & $4.7^{+0.9}_{-1.2}$ & Column density \\ [0.2cm]
              & $\log(\xi_{\rm abs}$/erg~cm~s$^{-1}$)  & $1.8^{+0.6}_{-0.2}$ & Ionization state  \\ [0.2cm]
              & $\sigma_{\rm v}$(km~s$^{-1}$)  & $14147^{+3648}_{-2344}$  & Velocity broadening \\ [0.2cm] 
              & $\chi^2$/$\nu$ & 2353/2346 & Fit statistic  \\ [0.2cm]              
\hline
\end{tabular}}
\end{center}
\label{table1b}
\end{table*}

\begin{figure*}
\includegraphics[scale=0.32,angle=-0]{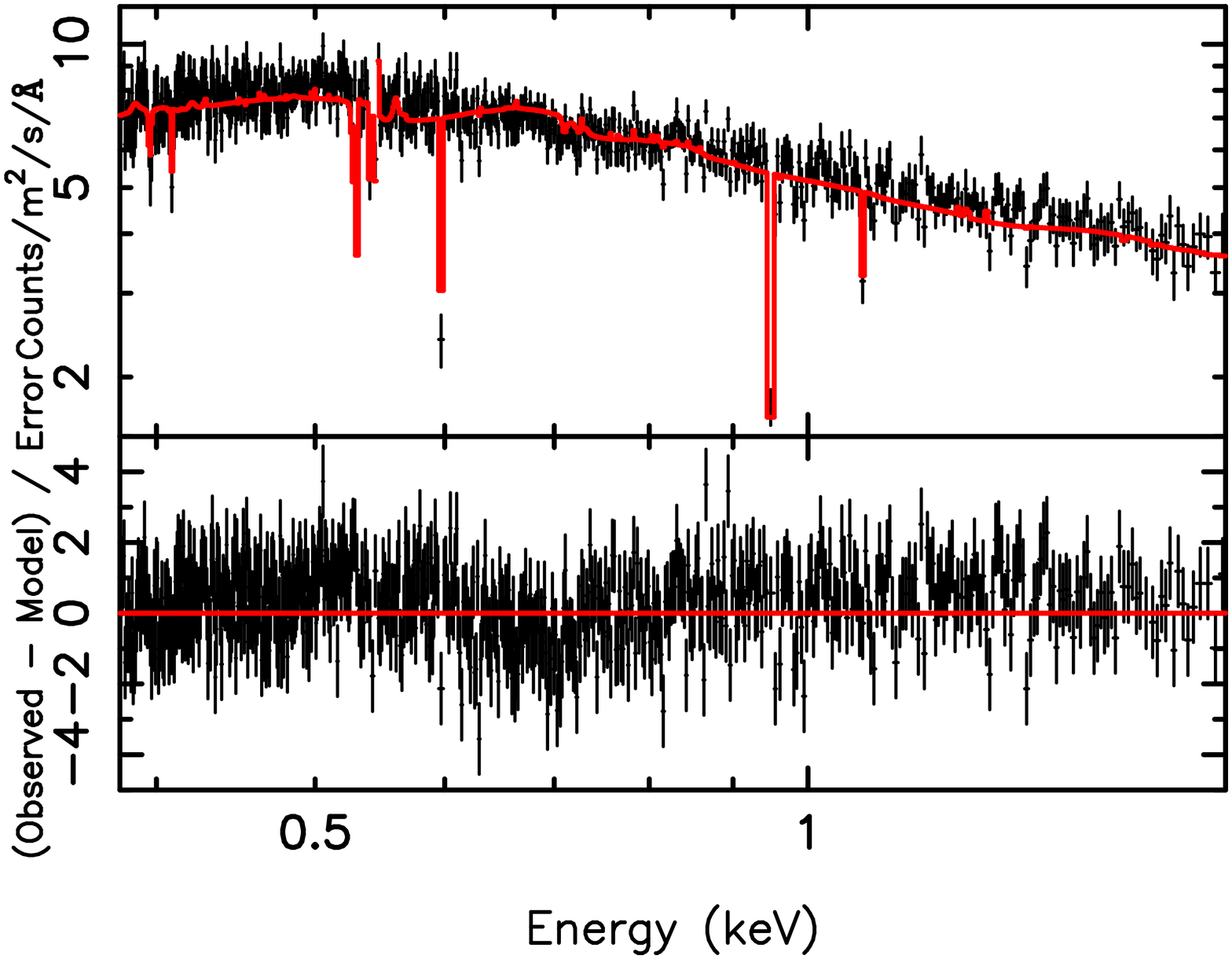}
\includegraphics[scale=0.32,angle=-0]{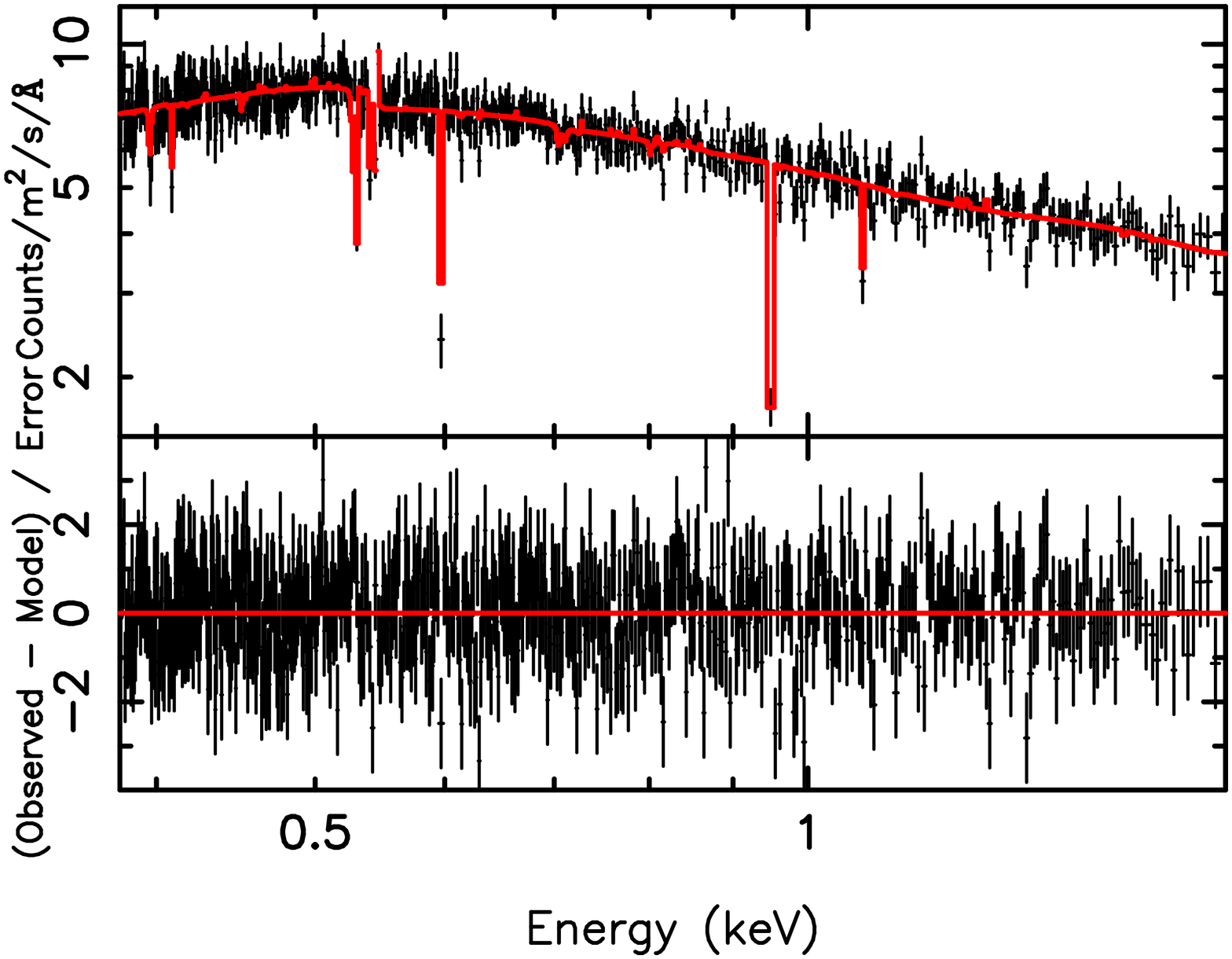}
\caption{Left: The 0.38$-$1.8\keV{} RGS spectrum, the spectral model, \textsc{hot$\times$reds$\times$(relxilld+xillverd+powerlaw)} and the residuals, showing a broad absorption feature at $\sim0.7$\keV{} in the observer's frame. Right: The 0.38$-$1.8\keV{} RGS spectrum, the best-fitting spectral model, \textsc{hot$\times$reds$\times$xabs$\times$(relxilld+xillverd+powerlaw)} and the residual plot. The data are binned up by a factor of 4 for plotting purposes only.}
\label{rgs}
\end{figure*}

\begin{table*}
 \centering
 \caption{The best-fitting model parameters for the joint fitting of \swift{}/XRT (0.3$-$6\keV{}), \xmm{}/EPIC-pn (0.3$-$10\keV{}) and \nustar{} (3$-$50\keV{}) spectra of Mrk~1044. Parameters with notations `(f)' and `$\ast$' indicate fixed and tied values, respectively.}
\begin{center}
\scalebox{0.95}{%
\begin{tabular}{ccccccccc}
\hline
Component&Parameter & \xmm{}/EPIC-pn & \swift{}/XRT  &  \nustar{} \\
& &  (0.3$-$10\keV{})& (0.3$-$6\keV{}) & (3$-$50\keV{}) \\
\hline 
Galactic absorption (\textsc{tbabs}) & $N_{\rm H}$(10$^{20}$~cm$^{-2}$) & $4.6^{+0.2}_{-0.2}$ & $4.6^{\ast}$ & $4.6^{\ast}$ \\[0.2cm]
Intrinsic absorption (\textsc{gabs}) & $E_{\rm abs}$(eV) & $707.2^{+7.1}_{-7.0}$ & $707.2^{\ast}$ & $707.2^{\ast}$ \\ [0.2cm]
             & $\sigma_{\rm abs}$(eV)  & $77.9^{+9.4}_{-8.3}$  & $77.9^{\ast}$ & $77.9^{\ast}$ \\ [0.2cm]             
       & $\tau_{\rm abs}$(10$^{-2}$)  & $1.9^{+0.2}_{-0.2}$ & $1.9^{\ast}$ & $1.9^{\ast}$ \\ [0.2cm]
Relativistic reflection (\textsc{relxilld}) & $q_{\rm in}$ &  $9.4^{+0.5}_{-0.7}$ & $9.4^{\ast}$    & $9.4^{\ast}$  \\ [0.2cm] 
                  & $q_{\rm out}$ &  $2.4^{+0.3}_{-0.2}$ & $2.4^{\ast}$   &  $2.4^{\ast}$  \\ [0.2cm]
                  & $r_{\rm br}$($r_{\rm g}$) &  $3.2^{+0.1}_{-0.1}$ & $3.2^{\ast}$ & $3.2^{\ast}$ \\ [0.2cm]
                 & $a$ &  $0.997^{+0.001}_{-0.016}$ & $0.997^{\ast}$ & $0.997^{\ast}$ \\ [0.2cm]  
                 & $\theta^{\circ}$  & $47.2^{+1.0}_{-2.5}$ &  $47.2^{\ast}$  & $47.2^{\ast}$ \\ [0.2cm]
                 & $r_{\rm in}$($r_{\rm g}$) &  $1.31^{+0.08}_{-0.05}$ & $1.31^{\ast}$ & $1.31^{\ast}$ \\ [0.2 cm]
                 & $r_{\rm out}$($r_{\rm g}$) & 1000(f) & 1000(f)  & 1000(f) \\ [0.2cm] 

                 & $\Gamma$  & $2.28^{\ast}$ & $2.31^{\ast}$ & $1.89^{\ast}$  \\ [0.2cm]       
                 & $A_{\rm Fe}$  & $2.3^{+0.1}_{-0.2}$& $2.3^{\ast}$ & $2.3^{\ast}$ \\ [0.2cm]    
                 & $\log(n_{\rm e}$/cm$^{\rm -3}$)  & $16.7^{+0.3}_{-0.7}$ & $16.7^{\ast}$ & $16.7^{\ast}$    \\ [0.2cm]    
                & $\xi_{\rm blur}$(erg~cm~s$^{-1}$) & $812^{+131}_{-64}$ & $81^{+108}_{-29}$ & $1980^{+235}_{-908}$ \\ [0.2cm]       
                & $N_{\rm blur}$(10$^{-4}$) & $1.2^{+0.1}_{-0.4}$ & $1.2^{\ast}$ & $1.2^{\ast}$\\ [0.2cm]       
Distant reflection (\textsc{xillverd}) & $\Gamma$ & $2.28^{\ast}$ & $2.31^{\ast}$ & $1.89^{\ast}$ \\ [0.2cm]       
                 & $A_{\rm Fe}$ & $2.3^{\ast}$ & $2.3^{\ast}$ & $2.3^{\ast}$ \\ [0.2cm]    
                 & $\log(n_{\rm e}$/cm$^{\rm -3}$) & 15(f) & $15^{\ast}$& $15^{\ast}$ \\ [0.2cm]    
                & $\xi_{\rm distant}$(erg~cm~s$^{-1}$)& 1(f) &1(f) & 1(f)\\ [0.2cm]       
                & $i^{\circ}$  & $60$(f) & $60^{\ast}$ & $60^{\ast}$ \\ [0.2cm]     
               & $N_{\rm distant}$(10$^{-6}$) & $2.3^{+0.7}_{-0.7}$ & $2.3^{\ast}$ & $2.3^{\ast}$ \\ [0.2cm]   
Incident continuum (\textsc{zpowerlw}) & $\Gamma$ & $2.28^{+0.01}_{-0.01}$ & $2.31^{+0.07}_{-0.06}$ & $1.89^{+0.14}_{-0.06}$ \\ [0.2cm]
       & $N_{\rm PL}$(10$^{-4}$) & $9.3^{+2.6}_{-1.2}$ & $11.0^{+1.9}_{-2.2}$ & $4.5^{+4.1}_{-4.4}$ \\ [0.2cm]  
     & $\chi^2$/$\nu$ & 735/694 & --- & ---  \\ [0.2cm]                   
\hline                        
\end{tabular}}
\end{center}
\label{table2}
\end{table*}

\begin{figure*}
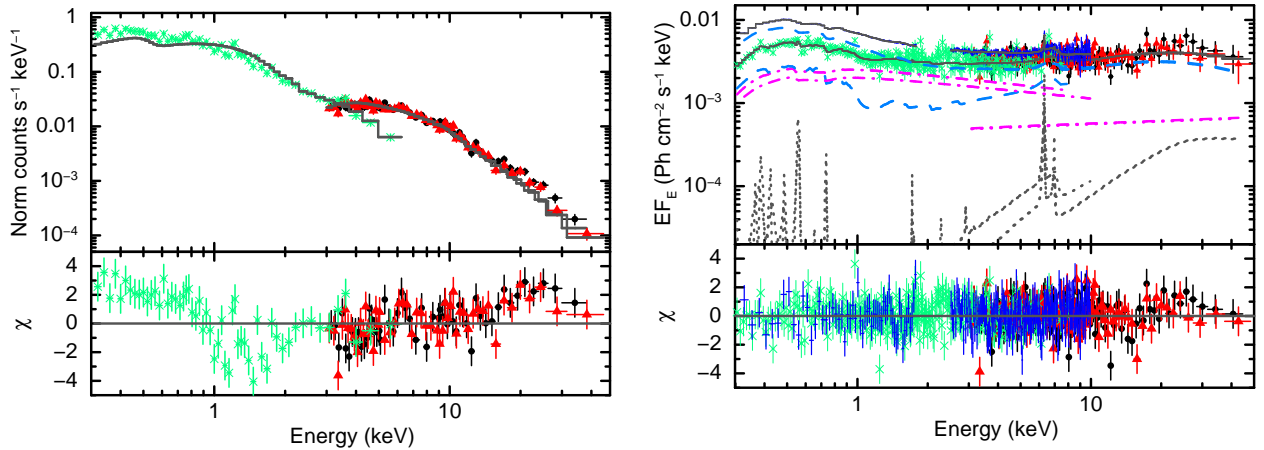

\includegraphics[scale=0.32,angle=-90]{fig7a.ps}
\includegraphics[scale=0.32,angle=-90]{fig7b.ps}
\caption{Left-hand panel: The \swift{}/XRT (0.3$-$6\keV{}: in green crosses), \nustar{}/FPMA (3$-$50\keV{}: in black circles) and FPMB (3$-$50\keV{}: in red triangles) spectral data, the primary power-law (\textsc{zpowerlw}: $\Gamma\sim2.13$) continuum model modified by the Galactic absorption (\textsc{tbabs}) fitted in the 0.3$-$50\keV{} band and the residuals as a function of energy. The residual plot shows a soft X-ray excess emission below $\sim1$\keV{}, a narrow emission line at $\sim6.4$\keV{} and a hard X-ray excess emission at around 15$-$30\keV{} energy range. The spectra are binned for clarity. Right-hand panel: The combined \swift{}/XRT (green crosses), \xmm{}/EPIC-pn (blue plus), \nustar{}/FPMA (black circles) and FPMB (red triangles) spectra, the best-fitting model, \textsc{tbabs$\times$gabs$\times$(relxilld+xillverd+zpowerlw)} and the residuals. The best-fitting model (in grey solid) consists of the following components: a primary power-law emission (in dash-dotted), a distant reflection (in dotted) for the 6.4\keV{} narrow Fe~K$_{\alpha}$ emission, and an ionized, high-density relativistic disc reflection (in dashed) responsible for the soft X-ray excess emission, broad Fe line and Compton hump.} 
\label{sw_nu} 
\end{figure*}

\subsection{The 0.3$-$10\keV{} EPIC-pn spectrum}
\label{full-pn}
To examine the presence of any excess emission in the soft X-ray band, we extrapolated the hard band (3$-$10\keV{}) best-fitting spectral model, \textsc{tbabs$\times$(relxilld$+$xillverd$+$zpowerlw)} down to 0.3\keV{}. Figure~\ref{pn_ldr} (left) shows the full band (0.3$-$10\keV{}) EPIC-pn spectral data, the extrapolated spectral model and the data-to-model ratio, which unveils an excess emission along with a broad absorption dip at $\sim0.7$\keV{} in the observed frame. For spectral analysis, we ignored the 1.8$-$2.5\keV{} band due to instrumental features present around the Si and Au detector edges at $\sim1.8$\keV{} and $\sim2.2$\keV{}, respectively (see e.g. \citealt{mar14,ma14}). Then we fitted the full band (0.3$-$10\keV{}) EPIC-pn spectrum with the hard band (3$-$10\keV{}) best-fitting model where the emissivity profile had a single power-law shape without any break. This resulted in a poor fit with $\chi^{2}$/d.o.f = 551.6/230. To fit the soft X-ray excess emission, we consider that the emissivity profile of the relativistic disc reflection follows a broken power-law shape: $\epsilon\propto r^{-q_{\rm in}}$ for $r<r_{\rm br}$ and $\epsilon\propto r^{-q_{\rm out}}$ for $r>r_{\rm br}$ where $r_{\rm br}$, $q_{\rm in}$ and $q_{\rm out}$ are the break radius, inner and outer emissivity indices, respectively. The fitting of the 0.3$-$10\keV{} EPIC-pn spectrum with the model, \textsc{tbabs$\times$(relxilld$+$xillverd$+$zpowerlw)}, where emissivity has a broken power-law shape, improved the fit statistics to $\chi^{2}$/d.o.f = 431/228 ($\Delta\chi^{2}$=$-$120.6 for 2 d.o.f). To model the absorption dip at $\sim0.7$\keV{}, we multiplied a Gaussian absorption line (\textsc{gabs}). The fitting of the 0.3$-$10\keV{} EPIC-pn spectral data with the high-density relativistic reflection plus constant density distant reflection as well as an intrinsic absorption [\textsc{tbabs$\times$gabs$\times$(relxilld+xillverd+zpowerlw)}] provided a statistically acceptable fit with $\chi^{2}$/d.o.f = 239.5/225. We also checked for the possible presence of warm absorbers and created an \textsc{xstar} table model \citep{ka01} with the default turbulent velocity of 300~km~s$^{-1}$, where we varied $\log\xi_{\rm wa}$ and $\log N_{\rm H}^{\rm wa}$ from $-5$ to $+5$ and $20$ to $25$, respectively. However, the modelling of the $\sim0.7$\keV{} absorption dip with the low-velocity warm absorber model(s) does not improve the fit statistics. An absorber with $\log\xi_{\rm wa}\sim0-1$ and $N_{\rm H}^{\rm wa}\sim5\times10^{20}$~cm$^{-2}$ yields minimal continuum curvature below $\sim0.6$\keV{} and an Fe~M UTA (unresolved transition array) feature close to $\sim0.8$\keV{}, which is too high for what is observed here. The lower ionization values, e.g., closer to $\log\xi_{\rm wa}\sim-1$ moves the Fe~M UTA close to $\sim0.7$\keV{} but adds moderate continuum curvature below $\sim0.5-0.6$\keV{} even for low column densities such as $N_{\rm H}^{\rm wa}\sim5\times10^{20}$~cm$^{-2}$. However, we did not see any extra continuum curvature below $\sim0.6$\keV{} even by fixing the Galactic neutral hydrogen column density at $N_{\rm H}=3.6\times10^{20}$\rm~cm$^{-2}$ \citep{wil13}. The centroid energies of the absorption line in the observed and rest frames are $\sim705.5$\ev{} and $\sim717.1$\ev{}, respectively. The $717.1$\ev{} absorption feature in the rest frame of the source can be due to either O~VII or O~VIII with corresponding outflow velocities of $\sim0.25c$ or $\sim0.1c$, where $c$ is the light velocity in vacuum. The interstellar hydrogen column density obtained from the 0.3$-$10\keV{} EPIC-pn spectral fitting is $N_{\rm H}=4.6^{+0.2}_{-0.3}\times10^{20}$\rm~cm$^{-2}$ where we considered the cross-sections and solar ISM abundances of \citet{aspl09}. The full band (0.3$-$10\keV{}) EPIC-pn spectrum, the best-fitting model, \textsc{tbabs$\times$gabs$\times$(relxilld+xillverd+zpowerlw)} and the deviations of the observed data from the best-fitting spectral model are shown in Fig.~\ref{pn_ldr} (right). In order to assure that the fitted parameter values are not stuck at any local minima, we performed a Markov Chain Monte Carlo (MCMC) analysis in \textsc{xspec}. We used the Goodman-Weare algorithm with 200 walkers and a total length of $10^{6}$. Figure~\ref{prob} shows the probability distributions of various parameters. To verify that the spectral parameters are not degenerate, we have shown the variation of different spectral parameters with the disc density and ionization parameter. Figure~\ref{cont} represents the contour plots between the disc density ($\log n_{\rm e}$) and other spectral parameters ($\Gamma$, $A_{\rm Fe}$, $\log\xi_{\rm blur}$, $q_{\rm in}$, $q_{\rm out}$, $r_{\rm br}$, $\theta^{\circ}$) and between disc ionization parameter ($\log\xi_{\rm blur}$) and two other spectral parameters ($r_{\rm in}$ and $q_{\rm in}$), which indicate that there is no degeneracy in the parameter space and the fitted parameters are independently constrained. The best-fitting spectral model parameters and their corresponding 90~per~cent confidence levels are summarized in Table~\ref{table1}. Thus, the \xmm{}/EPIC-pn spectral analysis indicates that the SE emission in Mrk~1044 results from the relativistic reflection off an ionized, high-density accretion disc with the ionization parameter of $\xi=909^{+90}_{-201}$~erg~cm~s$^{-1}$ and electron density of $\log(n_{\rm e}$/cm$^{\rm -3})=16.2^{+0.3}_{-0.1}$, respectively. We do not require any extra low-temperature Comptonization component to model the SE emission from the source. 
 
\subsection{The 0.38$-$1.8\keV{} RGS spectrum}
To confirm the presence of the $\sim0.7$\keV{} absorption line in the observed frame and also to detect any emission or absorption features, we have modelled the high-resolution RGS spectrum in \textsc{spex}~v.3.04. Since RGS data alone cannot constrain the continuum, we applied the full band (0.3$-$10\keV{}) EPIC-pn continuum model without any absorption component and redshift correction to the RGS data and multiplied a constant component to account for the cross-calibration uncertainties. We fixed all the continuum model parameters to the best-fitting EPIC-pn value. The EPIC-pn continuum model is then corrected for redshift and the Galactic ISM absorption with the models \textsc{reds} and \textsc{hot} in \textsc{spex}, respectively. The temperature of the \textsc{hot} model was fixed at $T=0.5$\ev{} \citep{pin13}, while the column density was set to vary freely. The fitting of the 0.38$-$1.8\keV{} RGS spectrum with the model, \textsc{hot$\times$reds$\times$(relxilld+xillverd+powerlaw)}, provided a $\chi^{2}$/d.o.f = 2411.5/2250 with a significant residual at $\sim0.7$\keV{} in the observed frame (Figure~\ref{rgs}, left). To model the absorption feature, we have used the photoionized absorption model (\textsc{xabs}) in \textsc{spex}~v.3.04. The \textsc{xabs} model calculates the absorption by a slab of material in photoionization equilibrium. The relevant parameters of the model are: column density ($N_{\rm H}^{\rm abs}$), line width ($\sigma_{\rm v}$), line-of-sight velocity ($v_{\rm out}$) and ionization parameter ($\log\xi_{\rm abs}$), where $\xi_{\rm abs}=\frac{L}{nr^{2}}$, $L$ is the source luminosity, $n$ is the hydrogen density and $r$ is the distance of the slab from the ionizing source. The modelling of the broad absorption dip with the \textsc{xabs} model improved the fit statistics to $\chi^{2}$/d.o.f = 2353/2346 ($\Delta \chi^{2}=-58.5$ for 4 d.o.f). The best-fitting parameters of the absorbers are summarized in Table~\ref{table1b}. The fitting of the rest-frame $\sim0.72$\keV{} absorption feature with a single absorber provides an outflow velocity of $v=(0.1\pm0.01)c$. If we fix the outflow velocity to $v=0.25c$ corresponding to the O~VII wind, then the fit statistics get worse by $\Delta\chi^{2}$=$8.1$ for 1 d.o.f. Thus our analysis prefers the slower ($0.1c$) O~VIII outflow over the faster ($0.25c$) O~VII outflow. The 0.38$-$1.8\keV{} RGS spectrum, the best-fitting spectral model, \textsc{hot$\times$reds$\times$xabs$\times$(relxilld+xillverd+powerlaw)} and the residuals are shown in Fig.~\ref{rgs} (right).

\subsection{Joint fitting of \swift{}, \xmm{} and \nustar{} spectra}
We jointly fitted the \swift{}/XRT, \xmm{}/EPIC-pn and \nustar{}/FPMA, FPMB spectra to obtain tighter constraints on the reflection continuum including the SE emission and investigate the presence of any hard X-ray excess emission above 10\keV{}. We tied all the parameters together between four data sets and multiplied a constant component to take care of the cross-normalization factors. This factor is kept fixed at 1 for the FPMA and varied for the FPMB, XRT and EPIC-pn spectral data. Initially, we fitted the quasi-simultaneous \swift{}/XRT and \nustar{}/FPMA, FPMB spectral data with a simple power-law (\textsc{zpowerlw}) model modified by the Galactic absorption (\textsc{tbabs}), which provided a poor fit with $\chi^{2}$/d.o.f = 601/329. The deviations of the observed data from the absorbed power-law model are shown in Figure~\ref{sw_nu} (left). The residuals show the presence of a soft X-ray excess emission below $\sim1$\keV{}, a narrow emission feature $\sim6.4$\keV{} and a hard X-ray excess emission in the energy range 15$-$30\keV{}, which most likely represents the Compton reflection hump as observed in many AGN (e.g. NGC~1365: \citealt{ri13}, MCG--6-30-15: \citealt{mar14}, 1H~070--495: \citealt{ka15}, Ark~120: \citealt{po18}). 

Then we applied the best-fitting EPIC-pn (0.3$-$10\keV{}) spectral model to the combined XRT, EPIC-pn, FPMA and FPMB spectral data sets. The joint fitting of the all four spectral data (0.3$-$50\keV{}) with the EPIC-pn spectral model, \textsc{tbabs$\times$gabs$\times$(relxilld+xillverd+zpowerlw)} provided a poor fit with $\chi^{2}$/d.o.f = 960/700. Since we are using multi-epoch observations, the spectral parameters might undergo significant variability. First, we set the photon index to vary between the \swift{}/XRT, \xmm{}/EPIC-pn and \nustar{} data sets. This improved the fit statistics to $\chi^{2}$/d.o.f = 776/698 ($\Delta\chi^{2}$=$-184$ for 2 d.o.f). Then we set the normalization of the incident continuum to vary between the observations, which improved the fit statistics to $\chi^{2}$/d.o.f = 765/696 ($\Delta\chi^{2}$=$-11$ for 2 d.o.f). Another spectral parameter that might encounter significant variability is the ionization parameter for the relativistic disc reflection component, which is directly proportional to the incident continuum flux. Therefore, we allow $\log\xi_{\rm blur}$ to vary, which provided an improvement in the fit statistics with $\chi^{2}$/d.o.f = 735/694 ($\Delta\chi^{2}$=$-30$ for 2 d.o.f) without any significant residuals. The cross-normalization factors obtained from the best-fitting model are $\sim1.02$ for FPMB, $\sim2.56$ for XRT and $\sim2.43$ for EPIC-pn, which is expected given the non-simultaneous observations over a long $\sim10$~year period. The best-fitting values for the electron density, inner radius, break radius and inclination angle of the disc are $\log(n_{\rm e}$/cm$^{\rm -3})=16.7^{+0.3}_{-0.7}$, $r_{\rm in}=1.31^{+0.08}_{-0.05}r_{\rm g}$, $r_{\rm br}=3.2^{+0.1}_{-0.1}r_{\rm g}$  and $\theta^{\circ}=47.2^{+1.0}_{-2.5}$, respectively. The inclusion of the data above 10\keV{} prefers moderately higher value for the disc density parameter. If we fix $\log(n_{\rm e}$/cm$^{\rm -3}$) to 15, then the broadband fit statistics get worse by $\Delta\chi^{2}$=$12.25$ for 1 d.o.f. An F-test indicates that the high-density disc model is preferred at the $\sim3.4\sigma$ confidence level compared to a low-density disc. Thus, the broadband (0.3$-$50\keV{}) spectral analysis suggests that the soft X-ray excess, highly ionized broad Fe line and Compton hump in Mrk~1044 is well explained by the relativistic disc reflection with a broken power-law emissivity profile of an ionized, high-density accretion disc. The \swift{}/XRT, \xmm{}/EPIC-pn, \nustar{}/FPMA and FPMB spectral data sets, the best-fitting model, \textsc{tbabs$\times$gabs$\times$(relxilld+xillverd+zpowerlw)} along with the components and the deviations of the broadband (0.3$-$50\keV{}) data from the best-fitting model are shown in Fig.~\ref{sw_nu} (right). The best-fitting broadband spectral model parameters are summarized in Table~\ref{table2}. If we replace the simple power-law (\textsc{zpowerlw}) model by a power-law with high-energy exponential cut-off (\textsc{cutoffpl}), then the estimated lower limit to the high-energy cut-off is $\sim128$\keV{}.

\begin{figure*}
\includegraphics[scale=0.32,angle=-0]{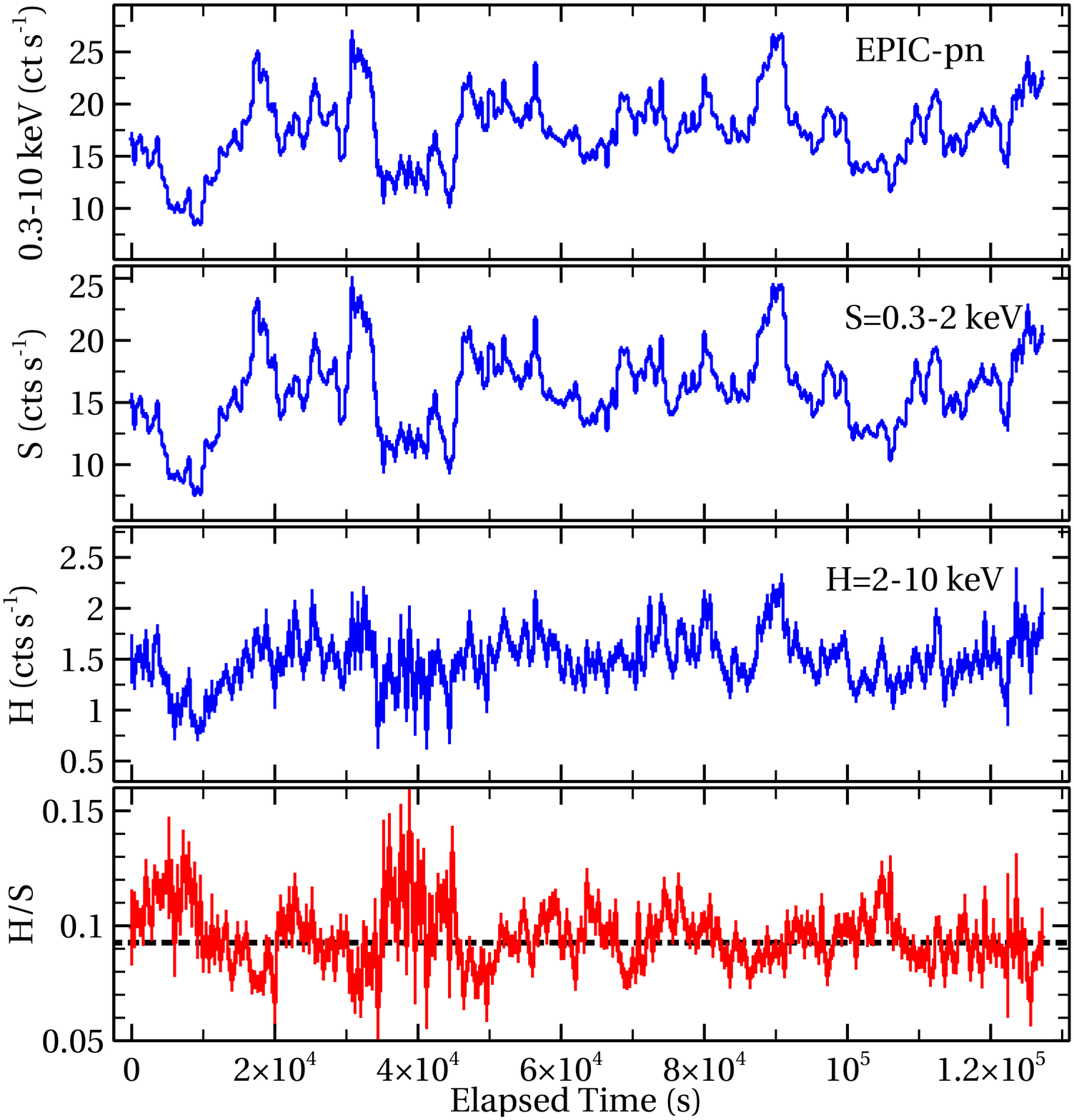}
\includegraphics[scale=0.32,angle=-0]{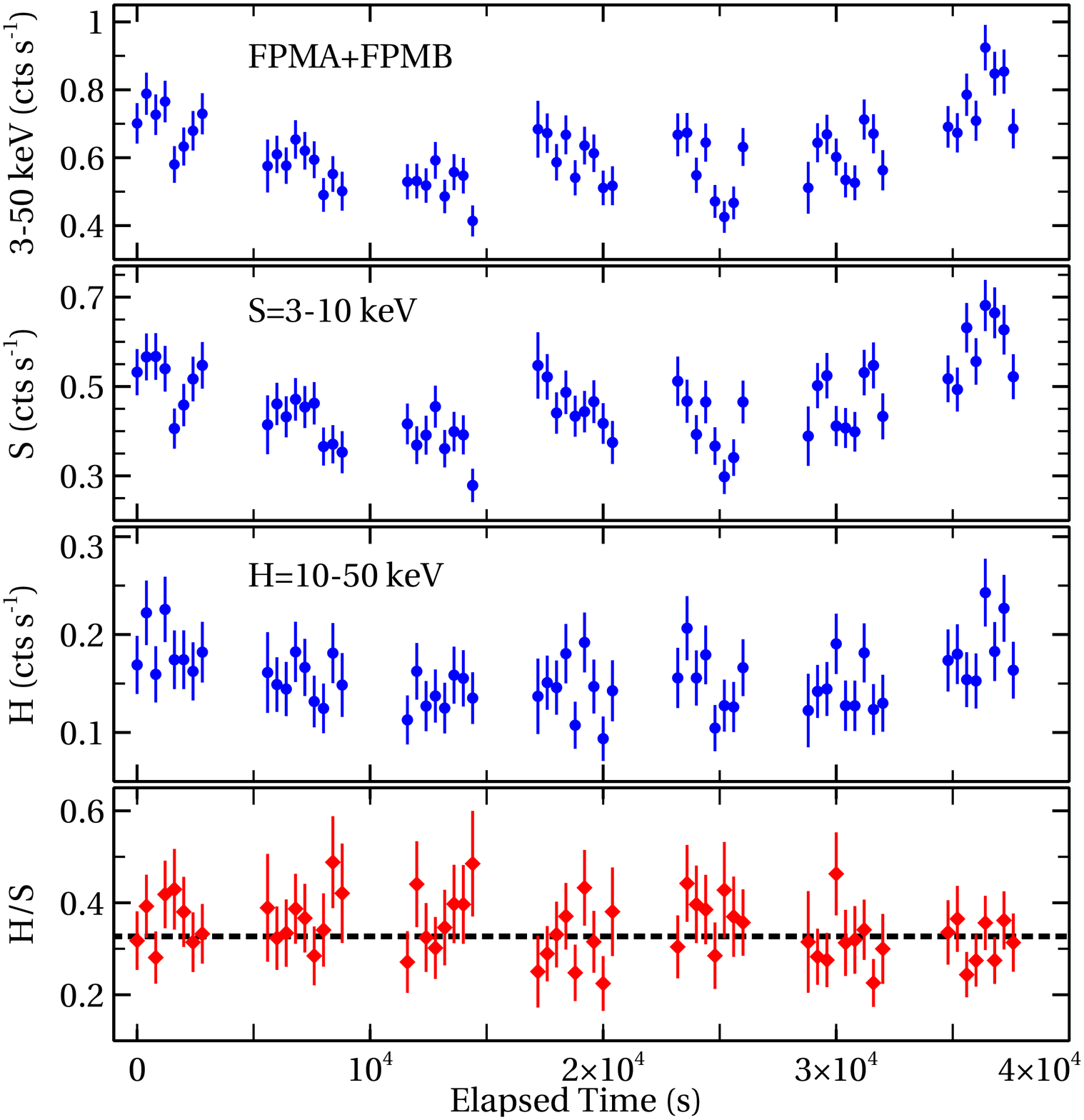}
\caption{Left: The background-subtracted, deadtime corrected \xmm{}/EPIC-pn light curves in the full (0.3$-$10\keV{}), soft (S=0.3$-$2\keV{}) and hard (H=2$-$10\keV{}) X-ray bands and the hardness ratio (H/S). Right: The background-subtracted \nustar{}/FPMA+FPMB light curves in three different energy bands: full (3$-$50\keV{}), soft (S=3$-$10\keV{}), hard (H=10$-$50\keV{}) and the corresponding hardness ratio (H/S) of Mrk~1044. The bin size used in all the panels is 400\s{}.}  
\label{lc}
\end{figure*}

\begin{figure}
\includegraphics[width=0.47\textwidth,angle=-0]{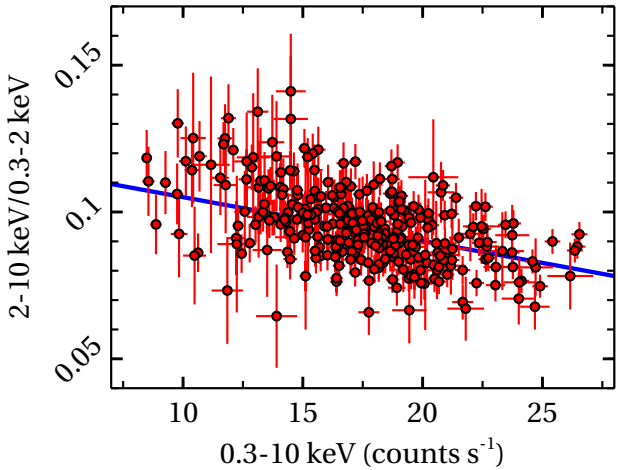}
\caption{The variation of the X-ray hardness ratio, H/S (S=0.3$-$2\keV{}, H=2$-$10\keV{}) as a function of the total (0.3$-$10\keV{}) X-ray count rate, representing a `softer-when-brighter' behaviour of Mrk~1044 as commonly observed in radio-quiet Seyfert~1 galaxies.}  
\label{hr_flux}
\end{figure}

\begin{figure*}
\includegraphics[width=0.45\textwidth,angle=-0]{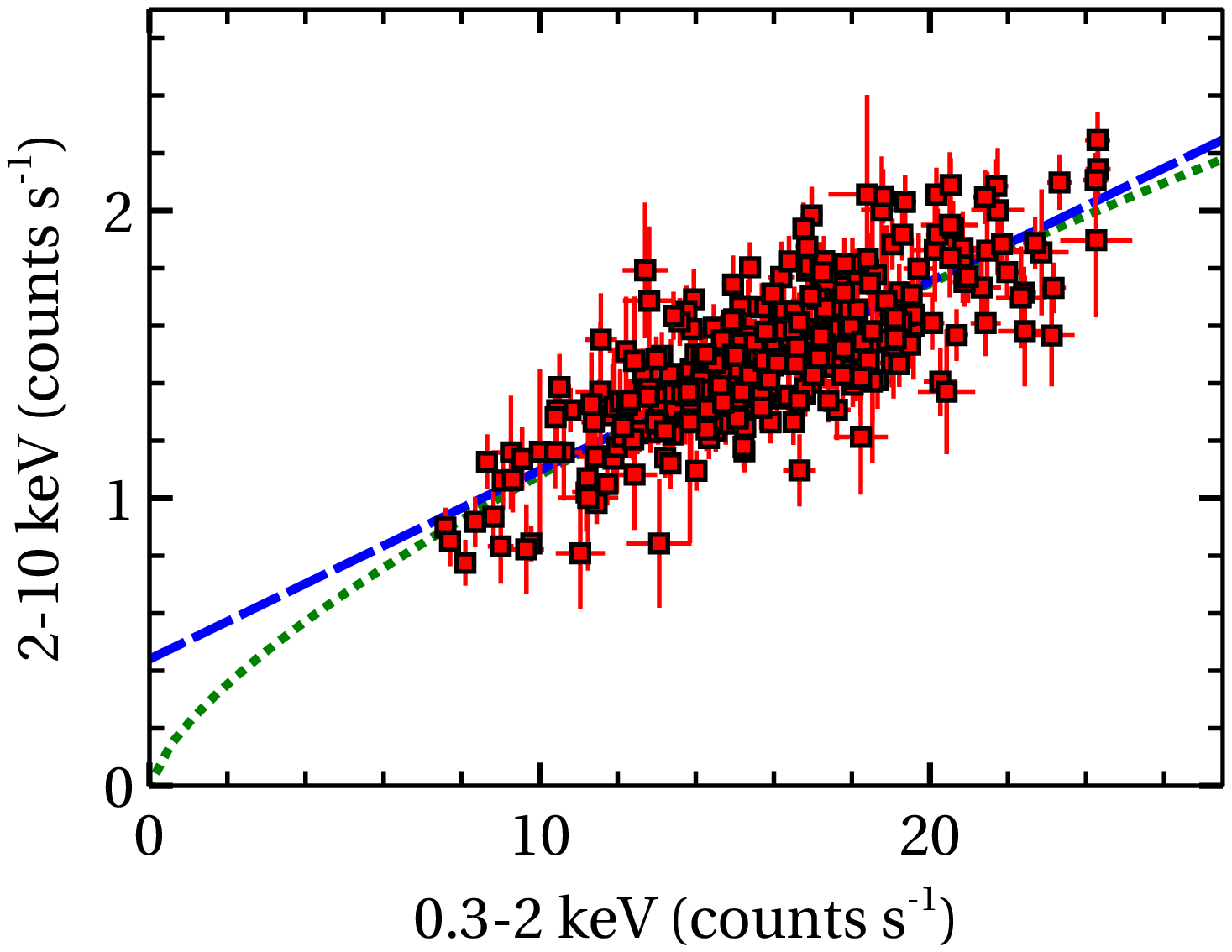}
\includegraphics[width=0.45\textwidth,angle=-0]{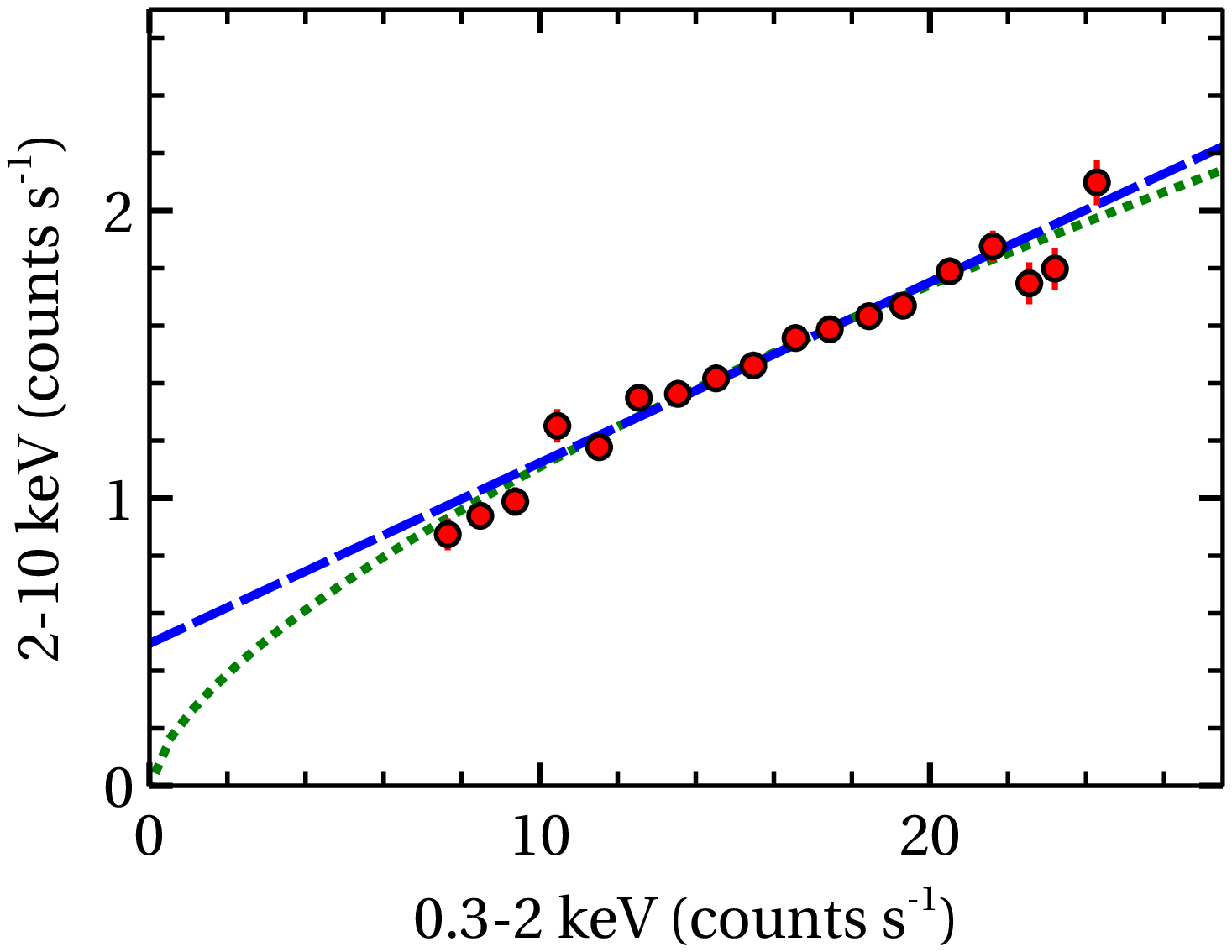}
\caption{The 0.3$-$2\keV{} vs. 2$-$10\keV{} flux$-$flux plot for Mrk~1044 (left). The time bin size used is 400\s{}. The binned flux$-$flux plot for Mrk~1044 (right). The dashed and dotted lines represent the linear (${\rm H}=m\times {\rm S}+c$) and power-law (${\rm H}=\alpha\times {\rm S}^{\beta}$) fits to the data.}  
\label{flux_flux}
\end{figure*}

\begin{figure*}
\includegraphics[width=0.47\textwidth,angle=-0]{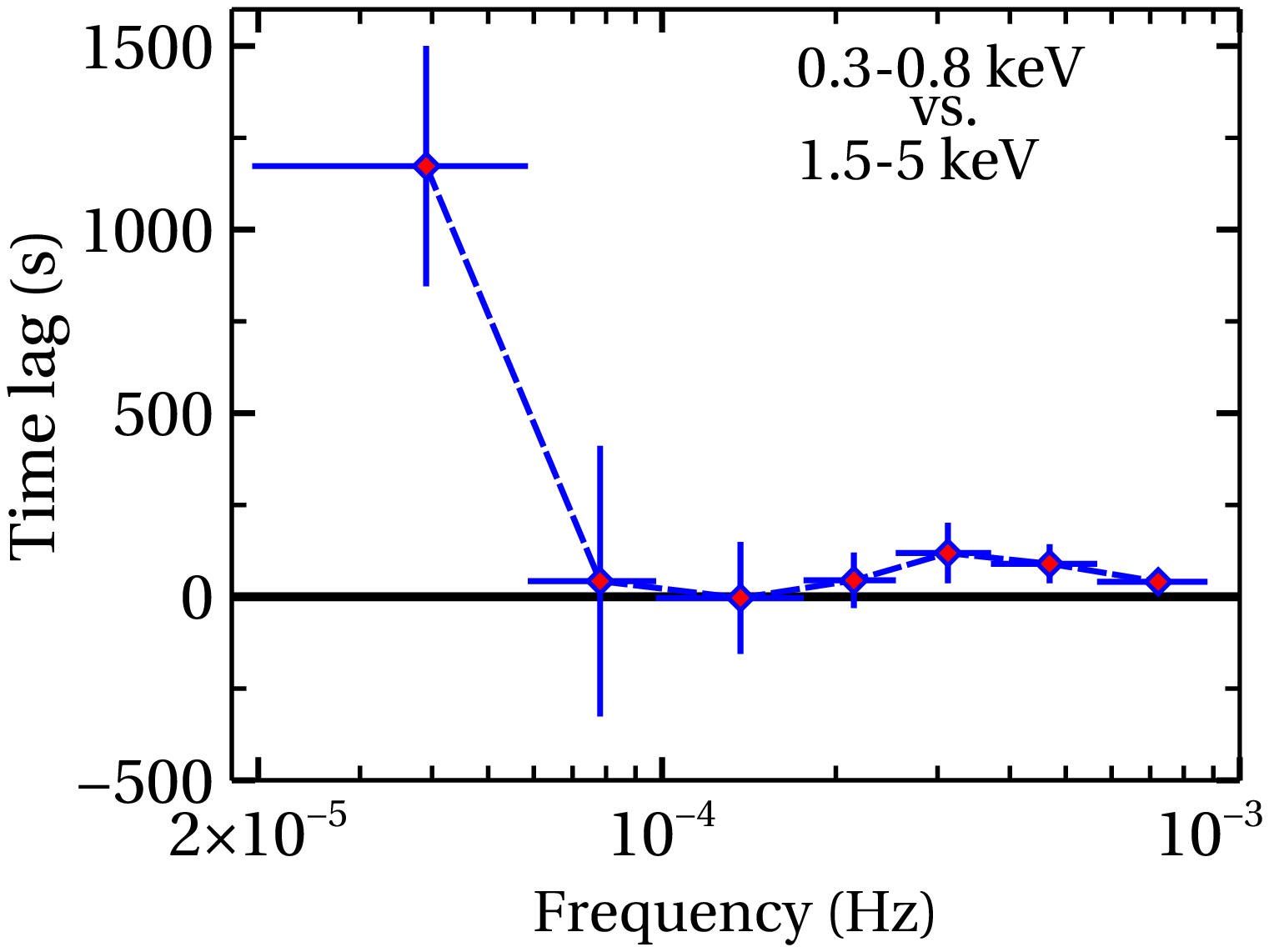}
\includegraphics[width=0.47\textwidth,angle=-0]{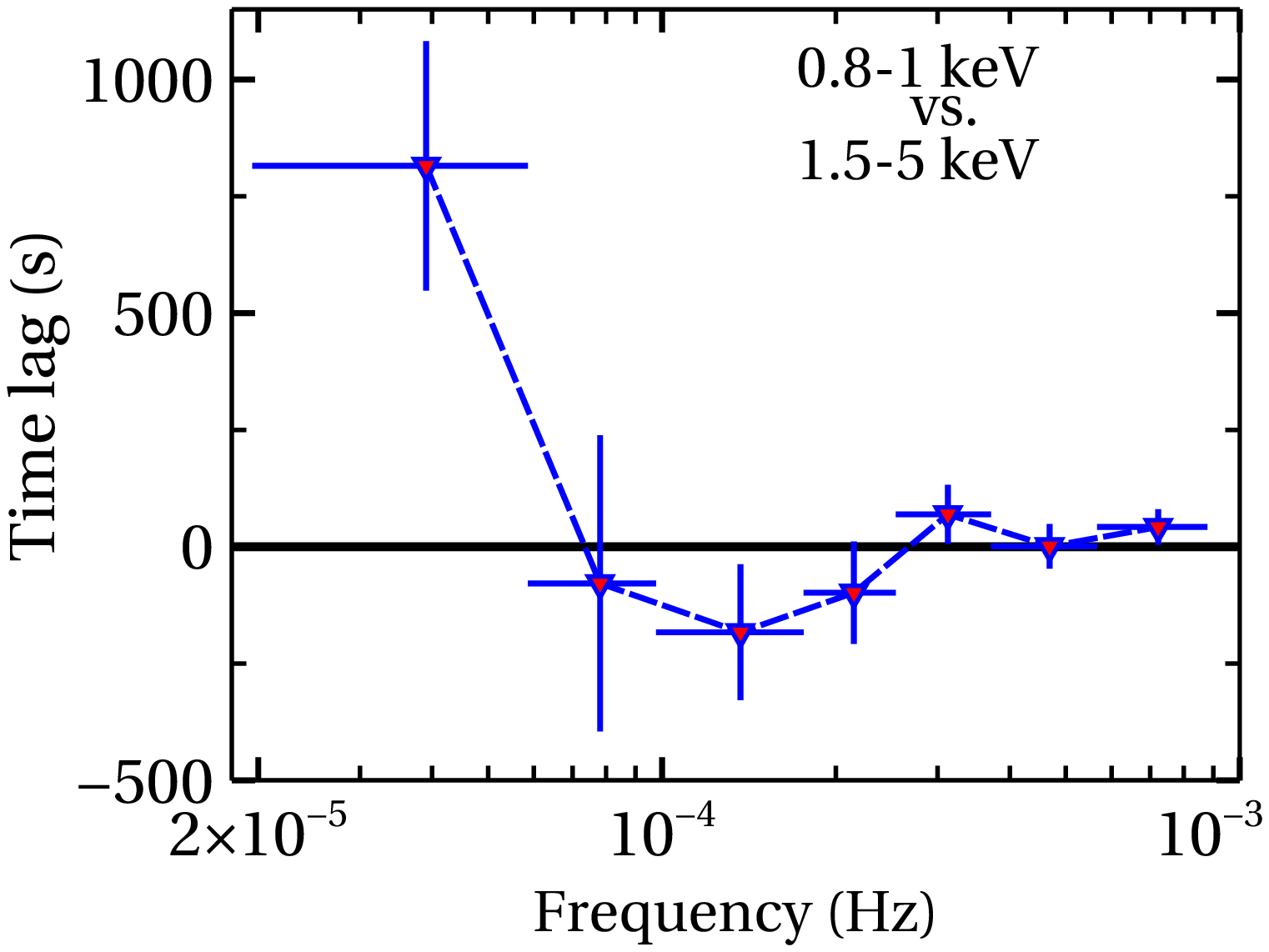}
\caption{Frequency-dependent time lags between the $E_{1}$ (0.3$-$0.8\keV{}) and $E_{3}$ (1.5$-$5\keV{}) light curves (left) and between the $E_{2}$ (0.8$-$1\keV{}) and $E_{3}$ (1.5$-$5\keV{}) light curves (right). Lags were calculated relative to the soft band ($E_{1}=0.3-0.8$\keV{}: left and $E_{2}=0.8-1$\keV{}: right), and positive lag implies that the hard band variations are delayed relative to the soft band variations.}
\label{lag_fre}
\end{figure*}

\begin{figure*}
\includegraphics[width=0.47\textwidth,angle=-0]{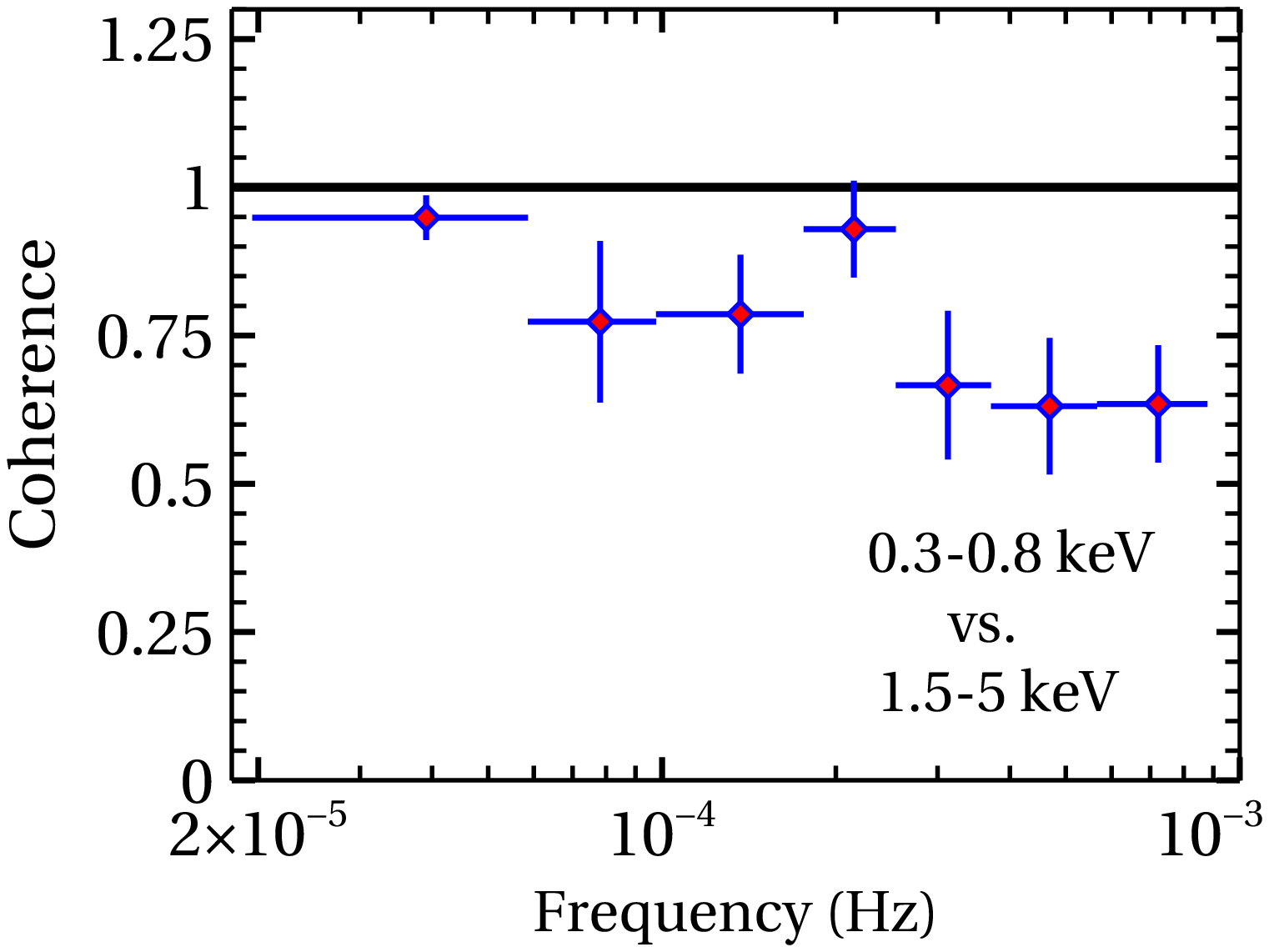}
\includegraphics[width=0.47\textwidth,angle=-0]{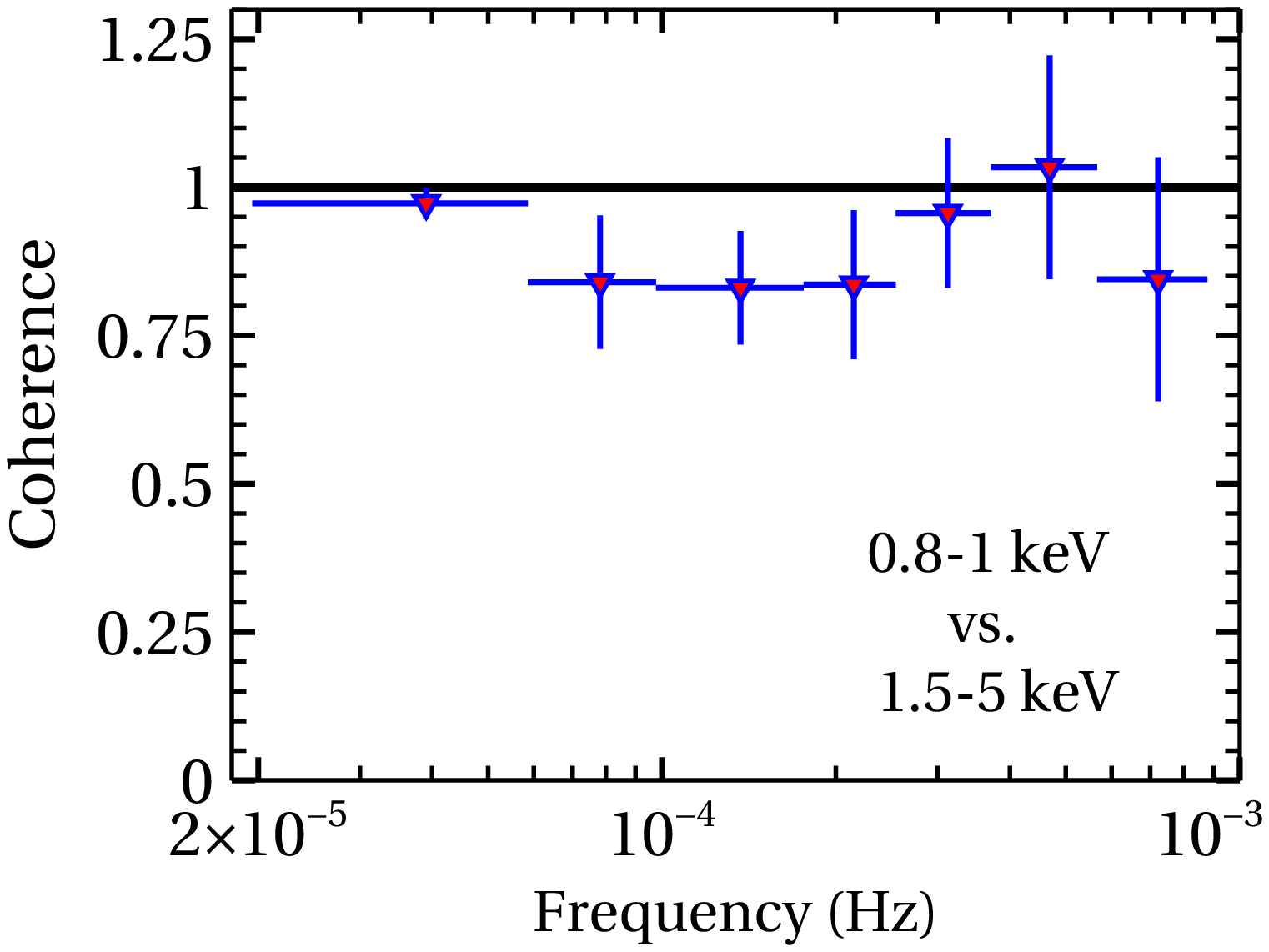}
\caption{Poisson noise subtracted coherence between 0.3$-$0.8\keV{} and 1.5$-$5\keV{} (left) and between 0.8$-$1\keV{} and 1.5$-$5\keV{} (right).}
\label{coh_fre}
\end{figure*}

\begin{figure*}
\includegraphics[width=0.47\textwidth,angle=-0]{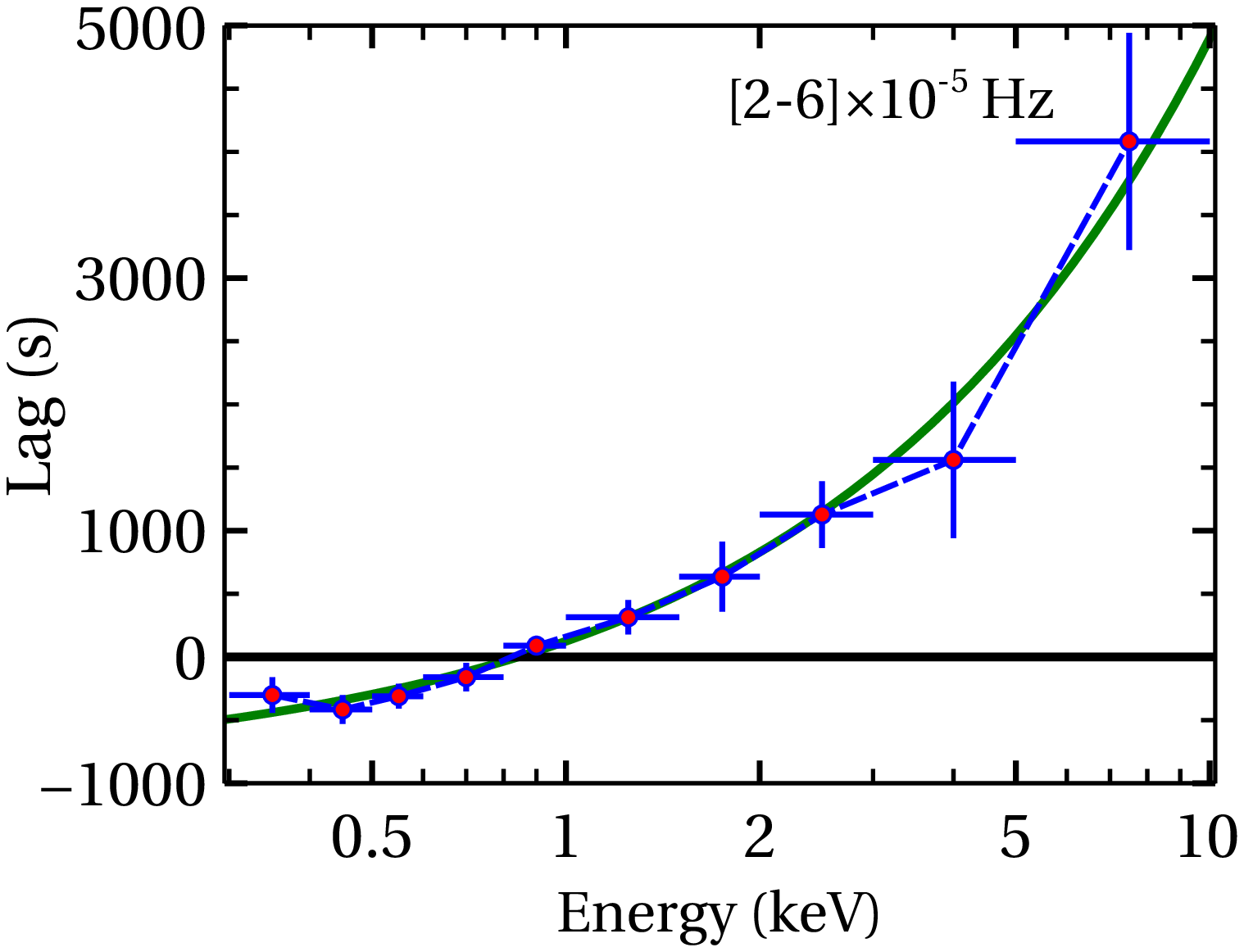}
\includegraphics[width=0.47\textwidth,angle=-0]{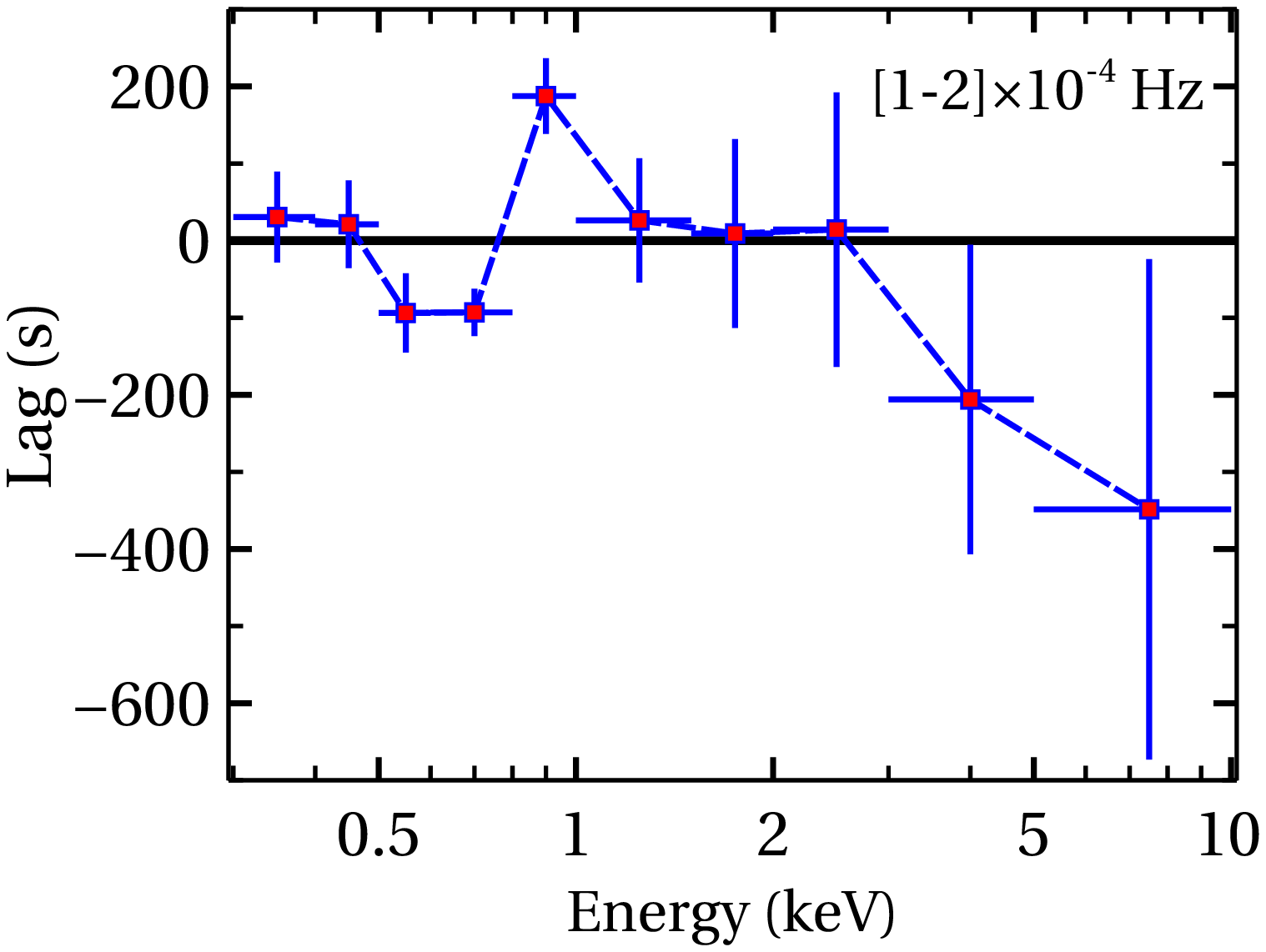}
\caption{Left: The lag-energy spectrum of Mrk~1044 in the lowest frequency range $\nu\sim[2-6]\times10^{-5}$\hz{}, where a hard lag is observed. The profile has a power-law like shape of the form $\tau(E)=1036.9\times E^{0.75}-912.2$\s{} and is shown as the solid, green line. Right: The lag-energy spectrum of Mrk~1044 in the high-frequency range $\nu\sim[1-2]\times10^{-4}$\hz{}, where a soft lag is detected. The high-frequency lag peaks at 0.8$-$1\keV{} and is very similar to the ratio of 0.3$-$10\keV{} data to the 3$-$10\keV{} best-fitting reflection model extrapolated down to 0.3\keV{}, as shown in the left, bottom panel of Fig.~\ref{pn_ldr}.}
\label{lag_E}
\end{figure*}

\begin{figure*}
\includegraphics[width=0.47\textwidth,angle=-0]{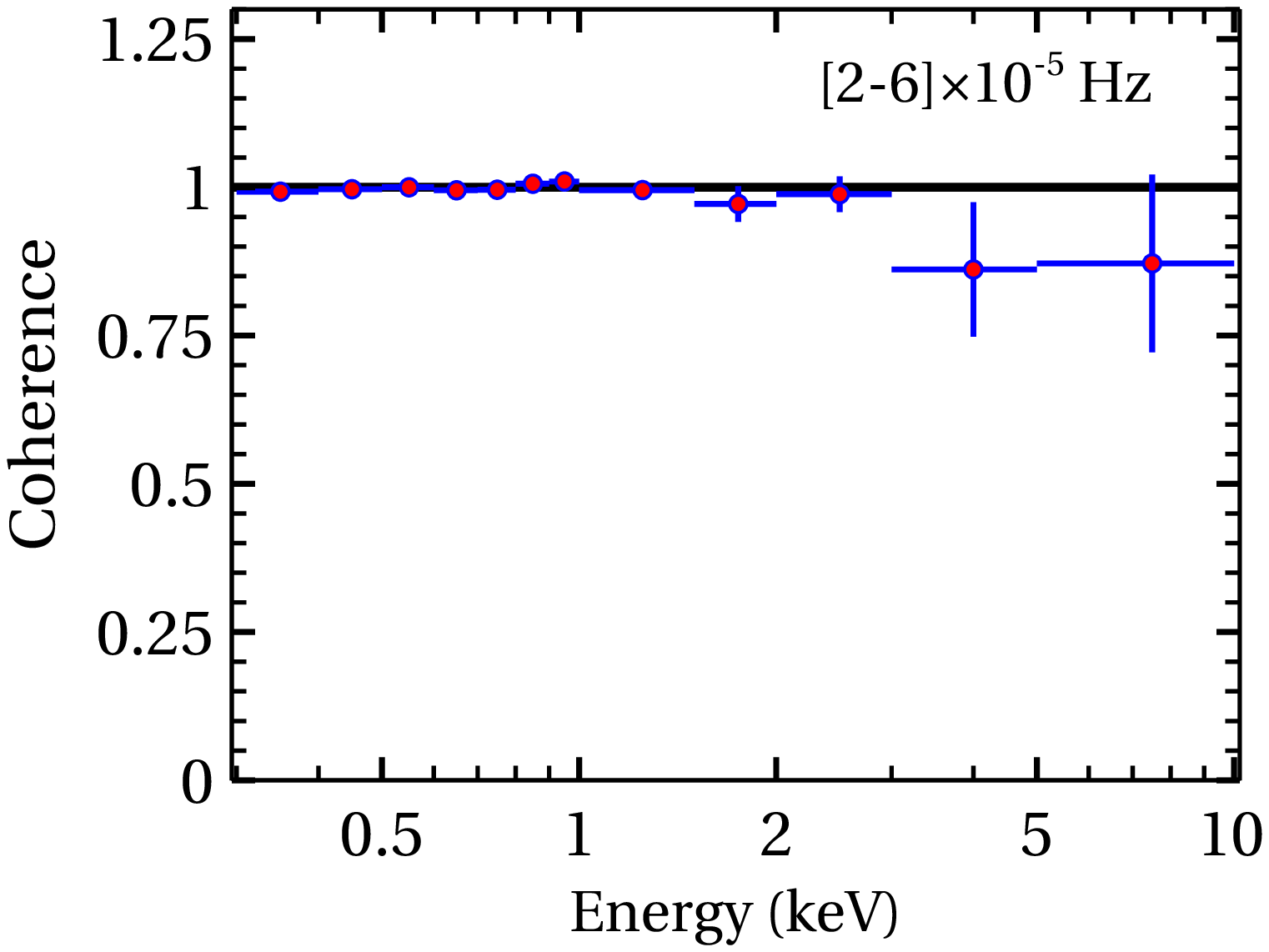}
\includegraphics[width=0.47\textwidth,angle=-0]{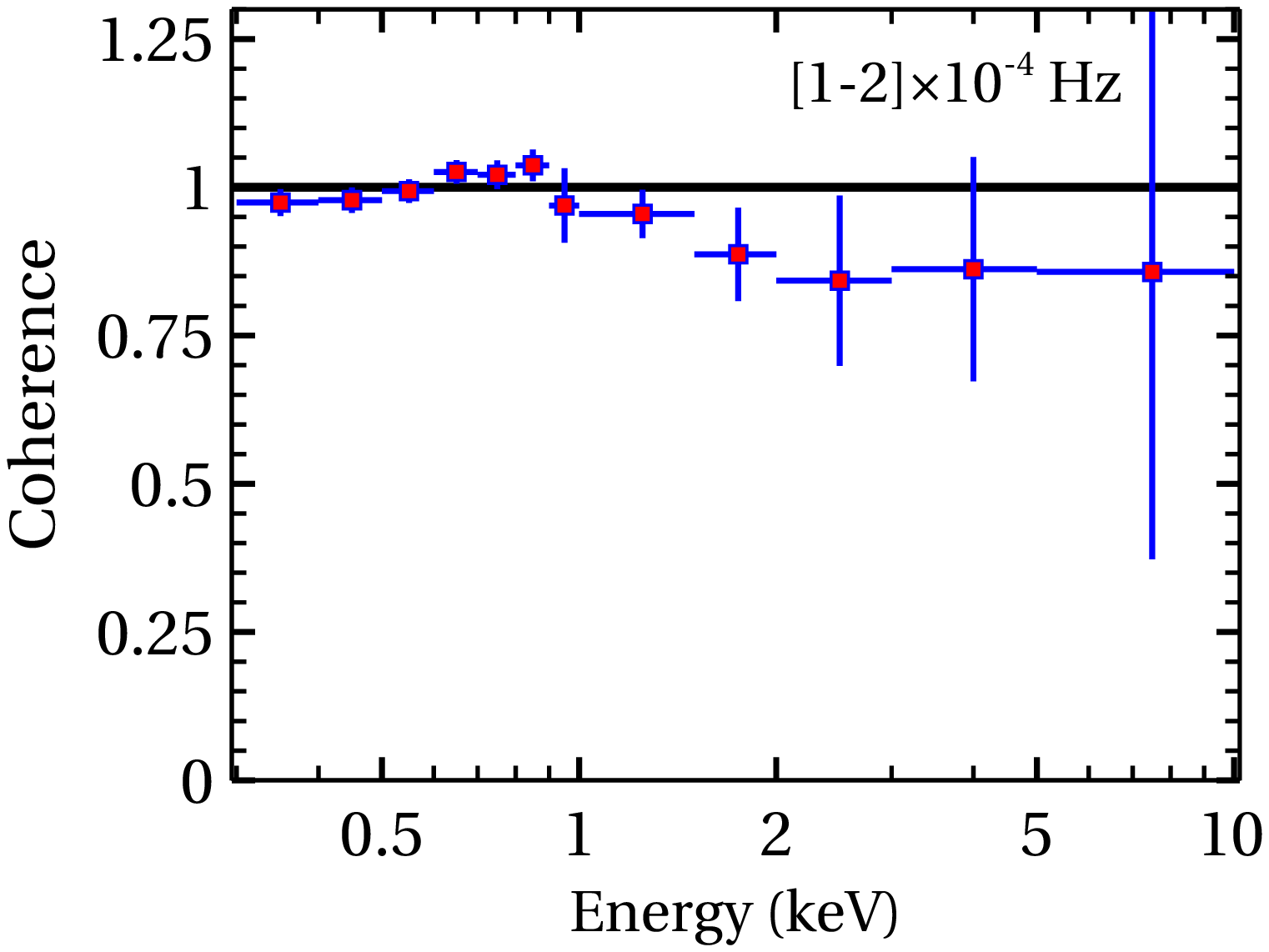}
\caption{Energy-dependent, noise corrected coherence in the lowest frequency range $\nu\sim[2-6]\times10^{-5}$\hz{} (left) and high-frequency range $\nu\sim[1-2]\times10^{-4}$\hz{} (right).}
\label{coh_E}
\end{figure*}

\begin{figure*}
\includegraphics[width=0.47\textwidth,angle=-0]{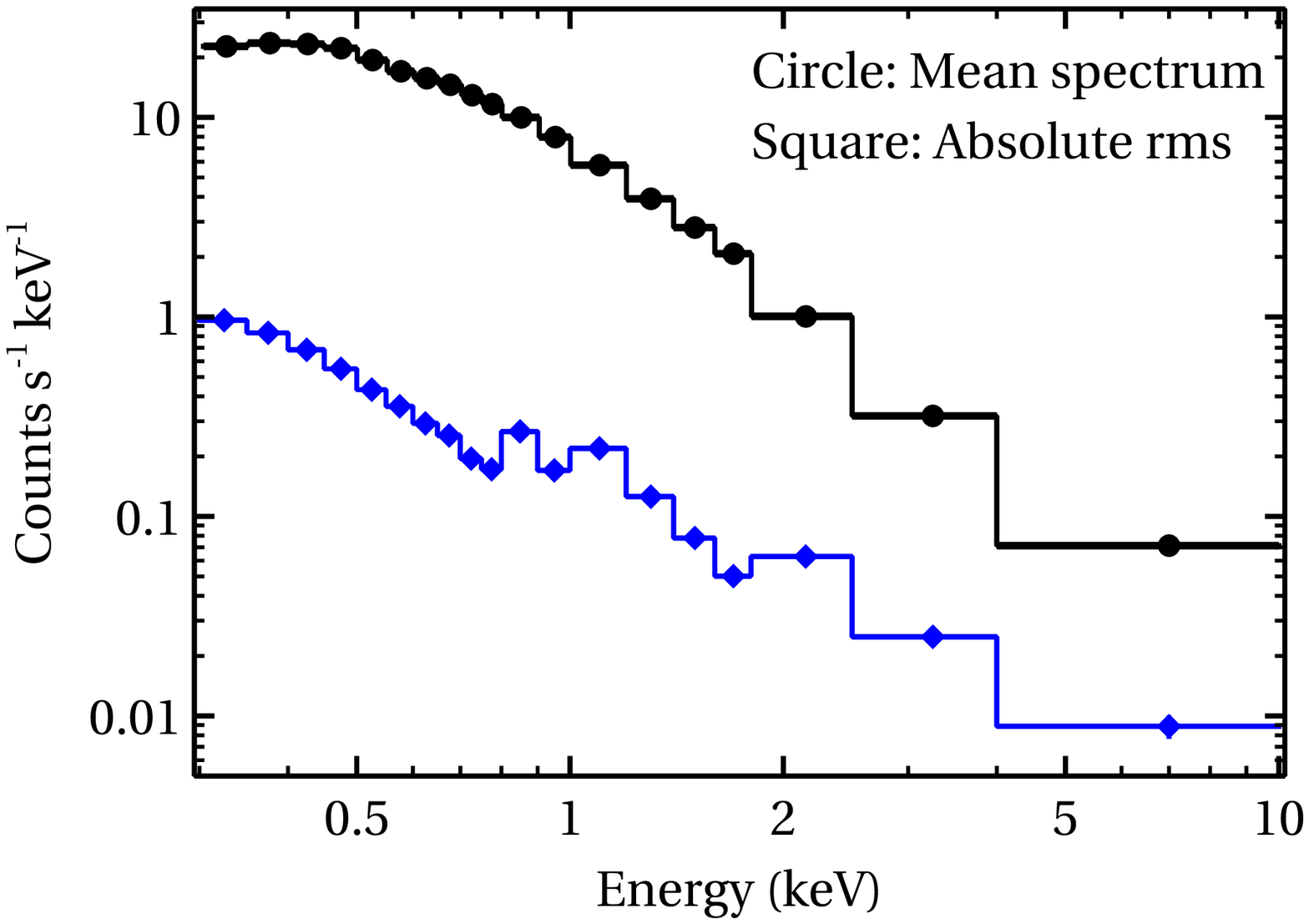}
\includegraphics[width=0.47\textwidth,angle=-0]{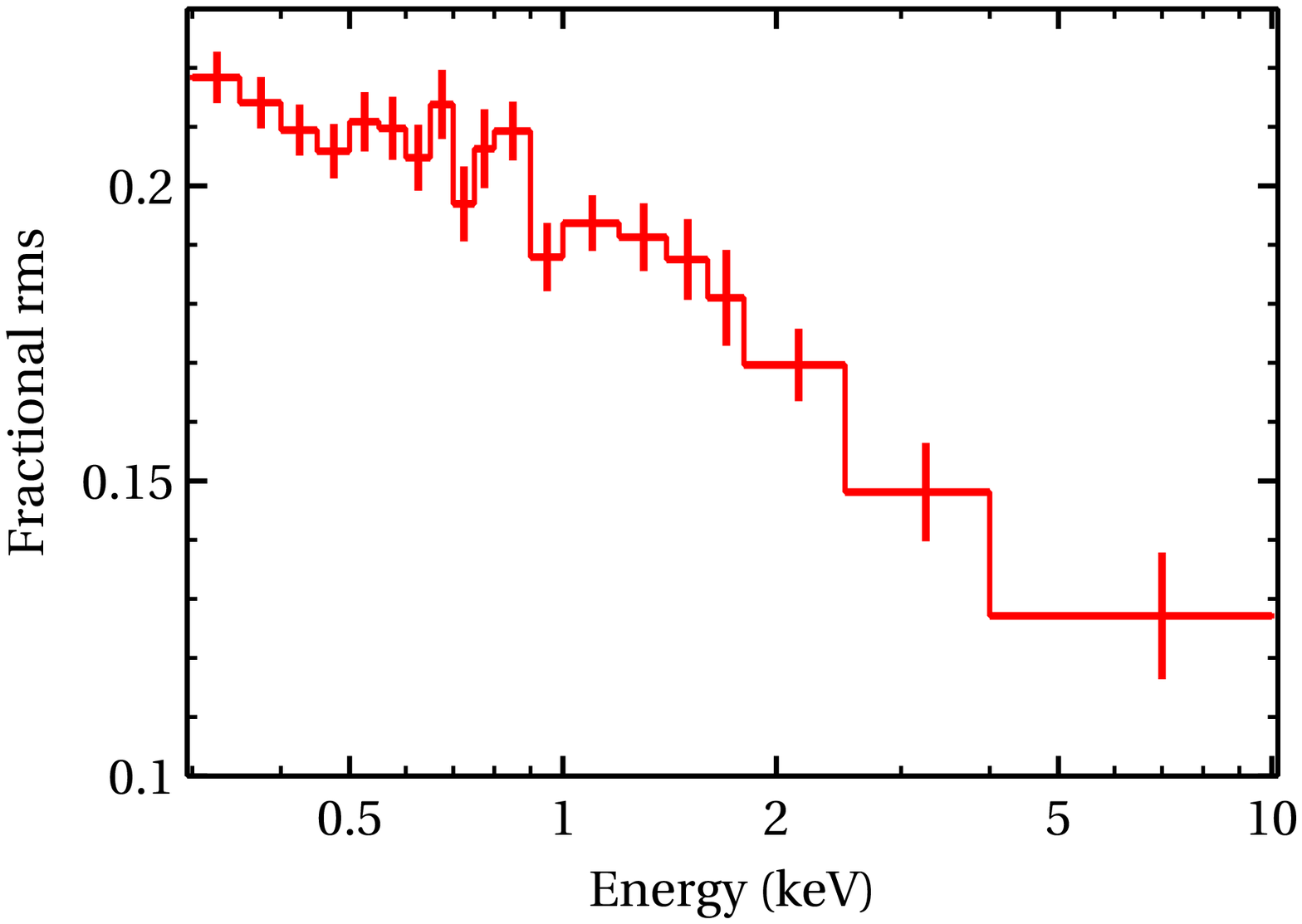}
\caption{Left: The 0.3$-$10\keV{} EPIC-pn mean (in circle) and frequency-averaged absolute rms (in square) spectra. Right: The 0.3$-$10\keV{} frequency-averaged fractional rms spectrum.}
\label{rms_data}
\end{figure*}

\begin{figure*}
\includegraphics[width=0.47\textwidth,angle=-0]{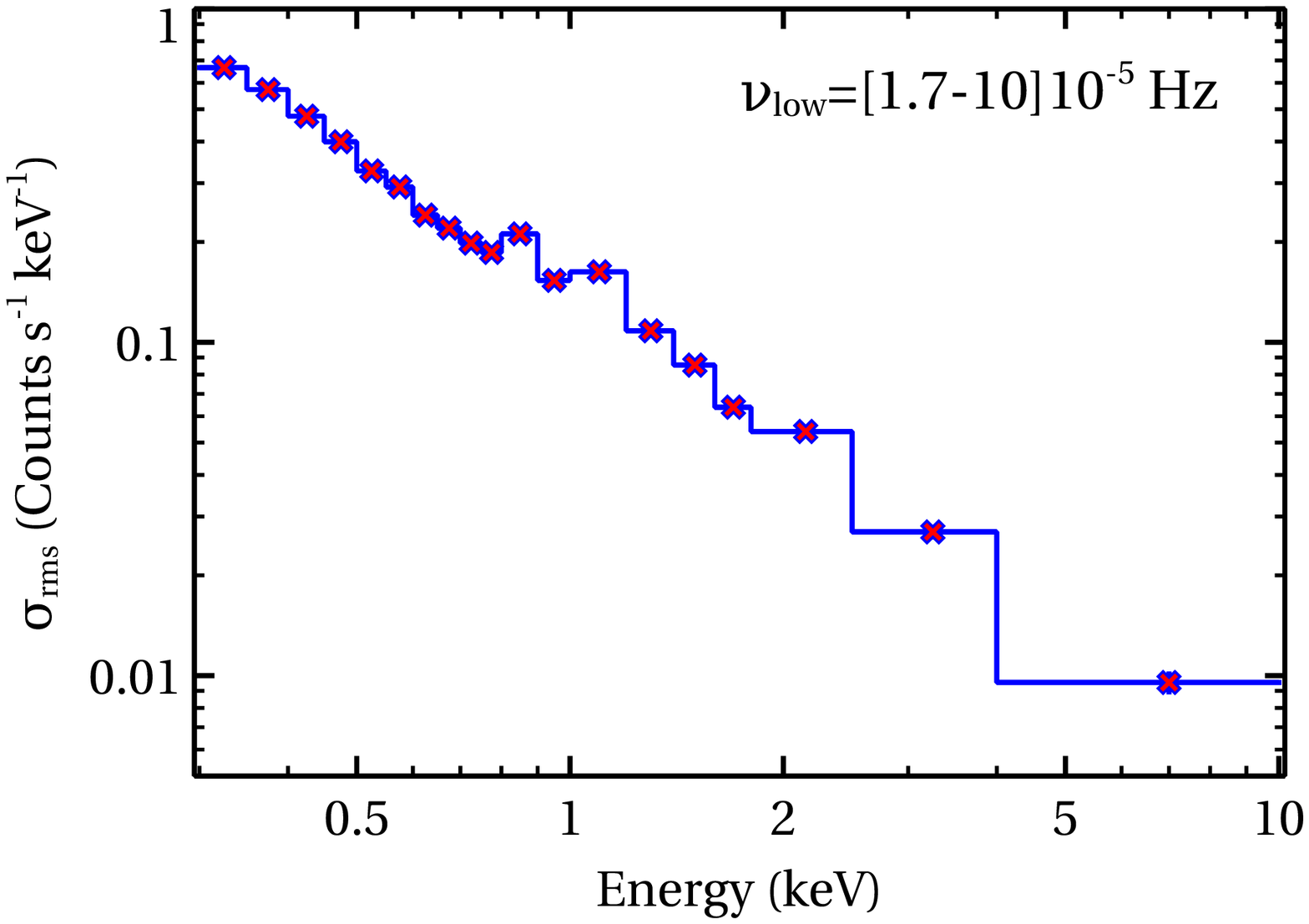}
\includegraphics[width=0.47\textwidth,angle=-0]{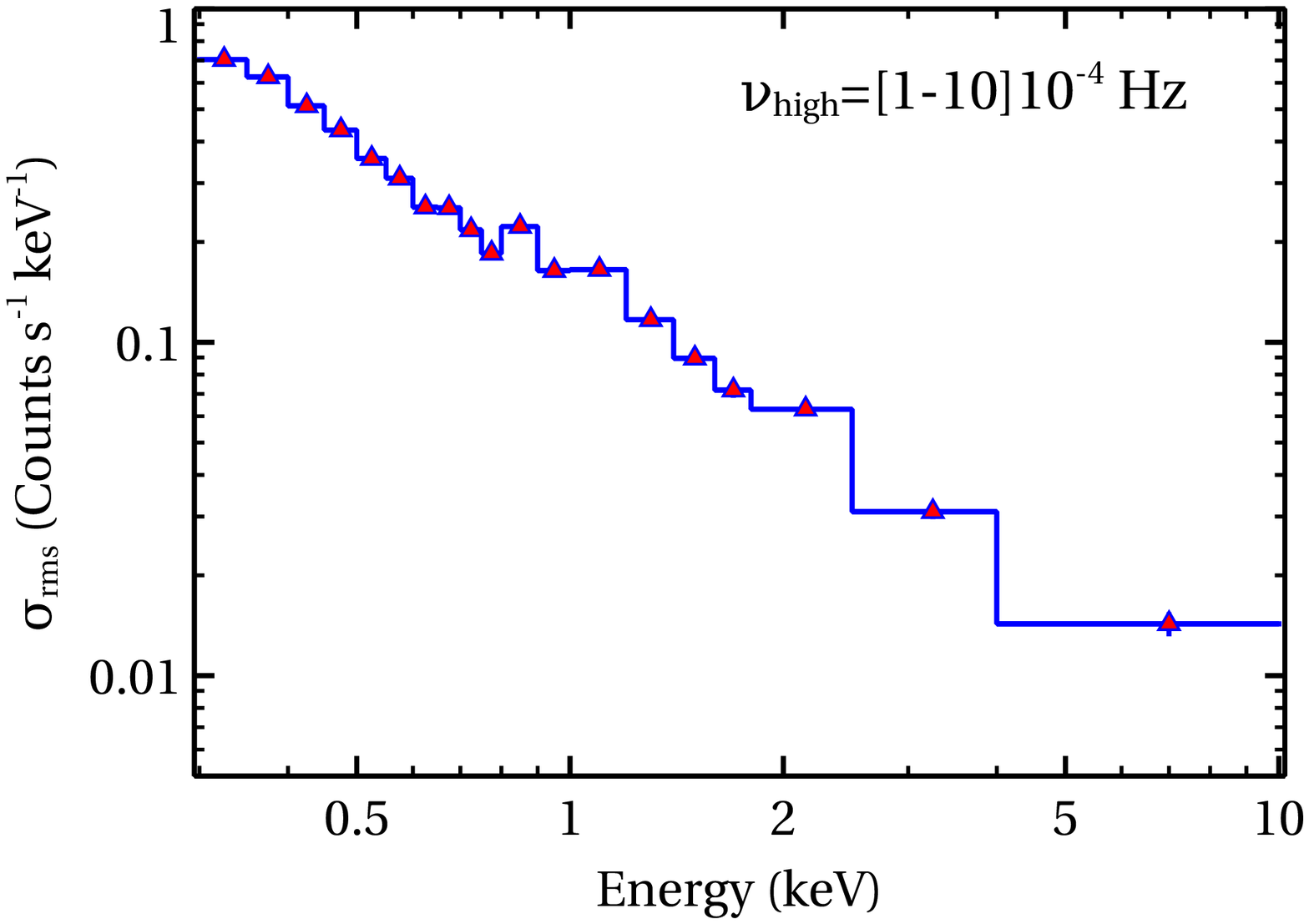}
\caption{The 0.3$-$10\keV{} absolute rms spectrum of Mrk~1044 in the low-frequency range, $\nu_{\rm low}\sim[1.7-10]\times10^{-5}$\hz{} (left) and high-frequency range, $\nu_{\rm high}\sim[1-10]\times10^{-4}$\hz{} (right). A hint of enhanced variability is found at $\sim0.85$\keV{} in both frequency bands.}
\label{rms_nu_resol}
\end{figure*}

\begin{figure*}
\includegraphics[width=0.47\textwidth,angle=-0]{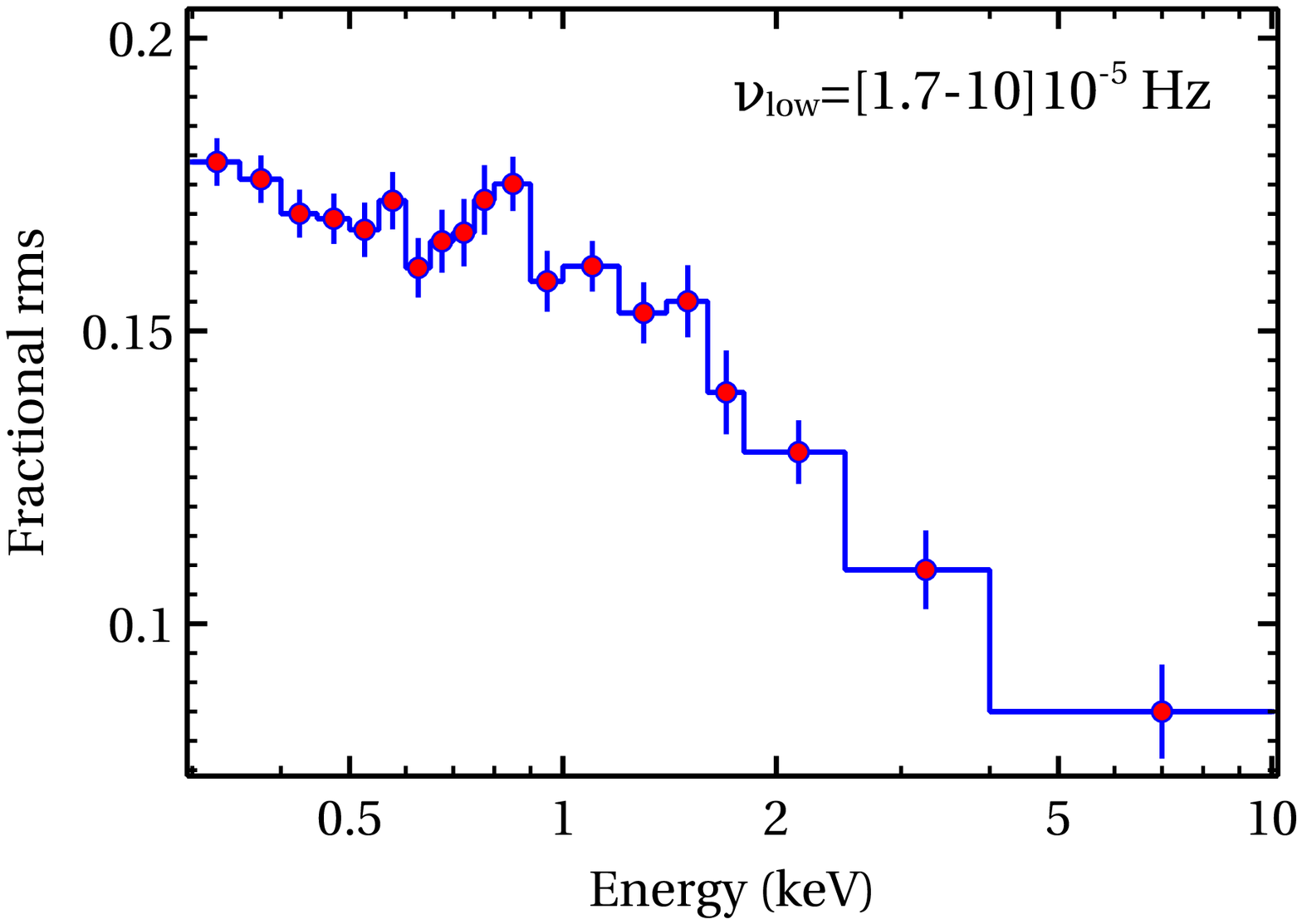}
\includegraphics[width=0.47\textwidth,angle=-0]{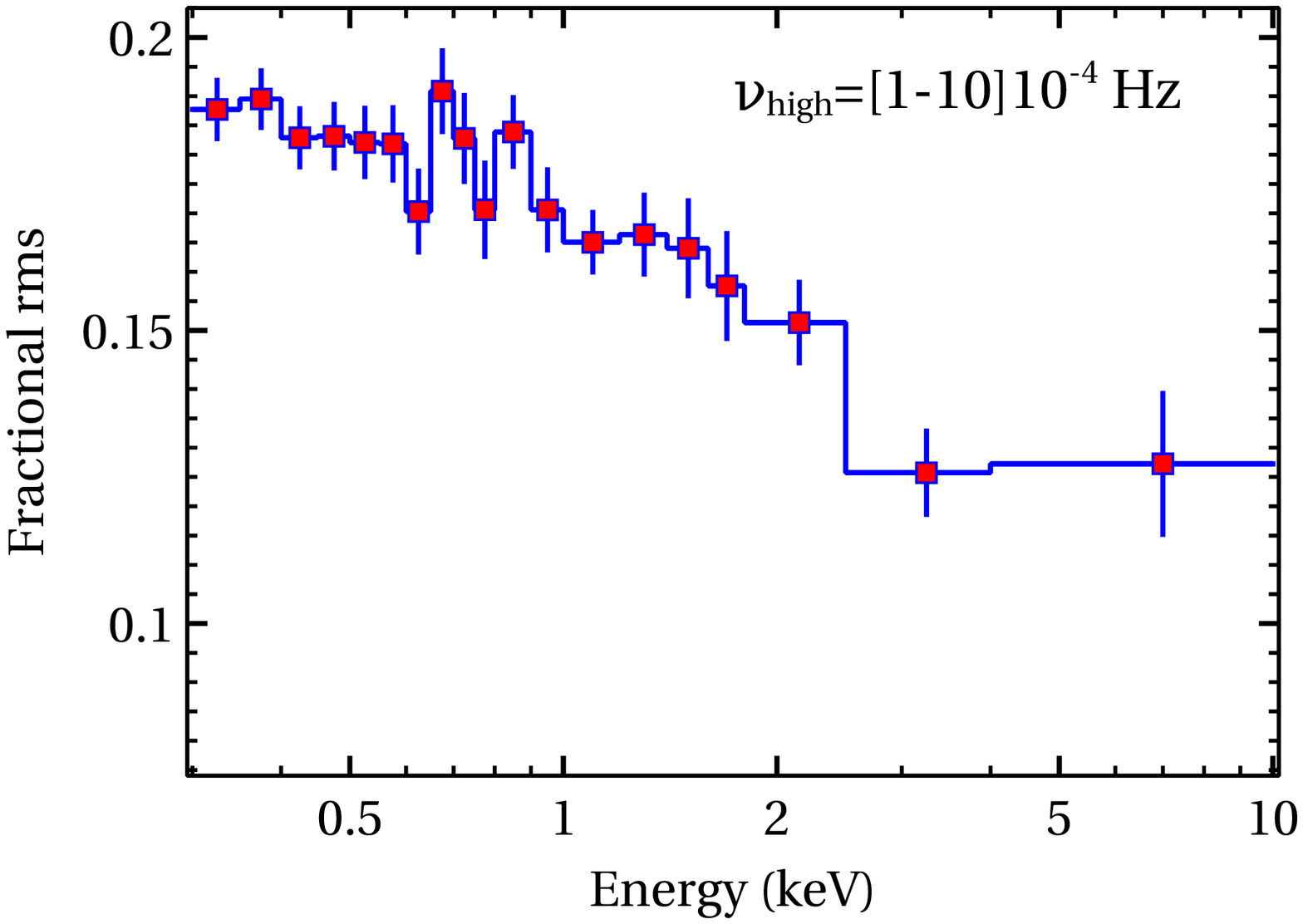}
\caption{The 0.3$-$10\keV{} fractional rms spectrum of Mrk~1044 in the low-frequency range, $\nu_{\rm low}\sim[1.7-10]\times10^{-5}$\hz{} (left) and high-frequency range, $\nu_{\rm high}\sim[1-10]\times10^{-4}$\hz{} (right).}
\label{fvar_nu_resol}
\end{figure*}

\begin{figure}
\includegraphics[width=0.47\textwidth,angle=-0]{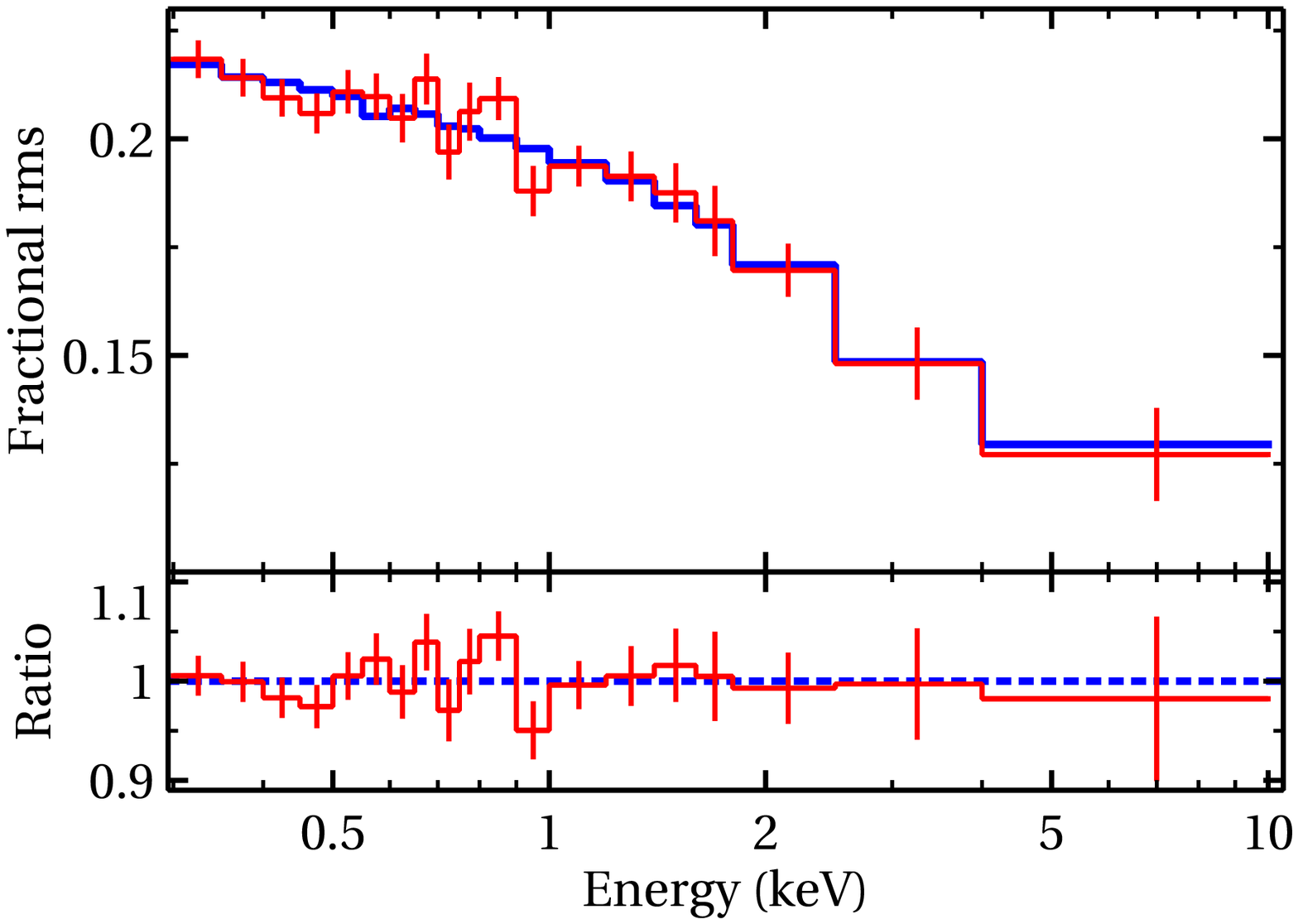}
\caption{The 0.3$-$10\keV{} frequency-averaged $F_{\rm var}$ spectrum, the best-fitting variability spectral model (in solid blue) and the data-to-model ratio. The best-fitting model has one constant distant reflection component and two either uncorrelated or moderately anti-correlated variable spectral components: an ionized high-density relativistic reflection continuum with variable normalization ($\Delta N_{\rm blur}$) and a direct power-law continuum with variable normalization ($\Delta N_{\rm PL}$) and spectral index ($\Delta \Gamma$), where $\Delta N_{\rm PL}$ and $\Delta \Gamma$ are positively correlated. The dotted line represents the data-to-model ratio of 1. Error-bars are 1$\sigma$.}
\label{rms_model}
\end{figure}

\section{Timing Analysis}
\label{sec:time}
To understand the time and energy dependence of the variability and the causal connection between the variable emission components, we explored the timing properties of Mrk~1044 using various model-independent approaches.

\subsection{Time series and hardness ratio}
Initially, we derived the background-subtracted, deadtime and vignetting corrected, full band (0.3$-$10\keV{}) EPIC-pn time series of Mrk~1044 with the time bin size of 400\s{}, as shown in the top, left panel of Figure~\ref{lc}. The source noticeably exhibits strong short-term variability. By applying the Scargle's Bayesian Blocks segmentation algorithm \citep{sc13} to the time series, we found that the X-ray emission from Mrk~1044 varied by $\sim50$~per~cent in about 800\s{}. The max-to-min amplitude variation of the source count rate is $\sim3.1$ on a timescale of $\sim22.8$~hr. To characterise the flux variations of the source, we computed the fractional rms variability, $F_{\rm var}$ and its $1\sigma$ error using the formula given in \citet{va03}. The estimated fractional rms amplitude in the full band (0.3$-$10\keV{}) is $F_{\rm var, 0.3-10}=(19.7\pm0.1$)~per~cent. We also investigated the energy dependence of variability by generating light curves in two different energy bands: soft (S=0.3$-$2\keV{}) and hard (H=2$-$10\keV{}), which are mostly dominated by the SE and primary emission, respectively. We have shown the soft and hard X-ray light curves of Mrk~1044 in Fig.~\ref{lc} (left). The variability trend in these two bands is found to be similar. However, the soft band is more variable than the hard band with the fractional rms variability of $F_{\rm var,0.3-2}=(20.3\pm0.1$)~per~cent and $F_{\rm var,2-10}=(15.5\pm0.5$)~per~cent, respectively which is indicative of the presence of multi-component variability. The max-to-min amplitude variations of the count rate in the 0.3$-$2\keV{} and 2$-$10\keV{} bands are $\sim3.2$ and $\sim2.9$ on timescales of $\sim22.4$~hr and  $\sim22.7$~hr, respectively. In the lower, left panel of Fig.~\ref{lc}, we have shown the temporal variations of the hardness ratio (HR), defined by H/S. The $\chi^{2}$ test revealed the presence of a significant variability in the source hardness with $\chi^{2}$/d.o.f=1028/318. The max-to-min amplitude variation of the hardness ratio is $\sim2.2$ on $\sim1.2$~hr timescales. Thus, we infer that Mrk~1044 showed a significant variability in flux as well as in spectral shape during the 2013 \xmm{} observation.

We also study the timing behaviour of the source in the 3$-$50\keV{} energy range during the \nustar{} observation in 2016. The full (3$-$50\keV{}), soft (3$-$10\keV{}) and hard (10$-$50\keV{}) X-ray, background-subtracted, combined FPMA and FPMB light curves of Mrk~1044 are shown in Fig.~\ref{lc} (right). The max-to-min amplitude variations in the 3$-$50\keV{}, 3$-$10\keV{} and 10$-$50\keV{} bands are $\sim2.2$, $\sim2.4$ and $\sim2.6$ on timescales of $\sim6.1$~hr, $\sim6.1$~hr and $\sim4.6$~hr, respectively. According to the $\chi^{2}$ test, the variability in the full (3$-$50\keV{}), soft (3$-$10\keV{}) and hard (10$-$50\keV{}) bands is statistically significant with $\chi^{2}$/d.o.f=200/58, 180/58 and 67/58, respectively. The fractional variability analysis confirms these results, indicating the presence of moderate variability with the amplitude of $F_{\rm var,3-50}=(14.4\pm1.3$)~per~cent, $F_{\rm var,3-10}=(15.1\pm1.5$)~per~cent and $F_{\rm var,10-50}=(7.8\pm4.8$)~per~cent in the full (3$-$50\keV{}), soft (3$-$10\keV{}) and hard (10$-$50\keV{}) bands, respectively. In the bottom, right panel of Fig.~\ref{lc}, we show the time series of the hardness ratio (HR=10$-$50\keV{}/3$-$10\keV{}). Although the source was variable in flux, the $\chi^{2}$ test revealed no significant variability in the hardness ($\chi^{2}$/d.o.f=42/58) during the 2016 \nustar{} observation.

\subsection{Hardness$-$flux and flux$-$flux analyses}
During the 2013 \xmm{} observation, the source was variable both in flux and spectral shape as shown in Fig.~\ref{lc} (left). To understand the interrelationship between the spectral shape and flux variations, we study the variation of the hardness ratio (HR=H/S: H=2$-$10\keV{} and S=0.3$-$2\keV{}) as a function of the total X-ray (0.3$-$10\keV{}) count rate. Figure~\ref{hr_flux} shows the hardness$-$intensity diagram of Mrk~1044, derived with the time bin size of 400\s{}. We found a decrease in source spectral hardness with the flux or `softer-when-brighter' behaviour of Mrk~1044, which is usually observed in radio-quiet Seyfert~1 galaxies (e.g. \citealt{mev03}, 1H~0707--495: \citealt{wi14}, Ark~120: \citealt{ma17,lo18}). To quantify the statistical significance of the observed `softer-when-brighter' trend of Mrk~1044, we estimated the Spearman rank correlation coefficient between the hardness ratio and X-ray count rate, which is $\sim-0.5$ with the null hypothesis probability of $p\sim1.2\times10^{-21}$. Although there is a lot of scatter in the diagram, a moderate anti-correlation between the hardness ratio and the total X-ray count rate is statistically significant.

We also studied the connection between the 0.3$-$2\keV{} and 2$-$10\keV{} band count rates using the flux$-$flux analysis which is a model-independent approach to study the spectral variability and was pioneered by \citet{ch01} and \citet{ta03}. We derived the flux$-$flux plot for Mrk~1044 with the time bins of 400\s{} (Figure~\ref{flux_flux}, left). We fitted the flux$-$flux plot with a linear model of the form ${\rm H}=m\times {\rm S}+c$, which provided a poor fit with $\chi^{2}$/d.o.f = 866/318, slope $m\sim0.07$ and a hard offset $c\sim0.44$~cts~s$^{-1}$. In order to investigate whether spectral pivoting of the primary power-law emission is a plausible cause for the observed variability, we fitted the data with a simple power-law model of the form ${\rm H}=\alpha\times {\rm S}^{\beta}$, which resulted in a statistically unacceptable fit with $\chi^{2}$/d.o.f = 862/318, $\alpha\sim0.7$ and $\beta\sim0.22$. The poor quality of the fitting could be caused by the presence of lots of scatter in the data. For greater clarity, we binned up the flux$-$flux plot in the flux domain. The binned flux$-$flux plot is shown in the right panel of Fig~\ref{flux_flux}. However, the fitting of the binned flux$-$flux plot with the linear and power-law models provided unacceptable fits with $\chi^{2}$/d.o.f = 37.4/16 and $\chi^{2}$/d.o.f = 28.9/16, respectively. To effectively distinguish between a linear versus a power-law fit, a wide range of flux is required. However, the EPIC-pn count rate spans only a factor of $\sim2-3$ for Mrk~1044. The error bars on the binned flux-flux points are too small to yield a good fit without adding some systematics. Another possible reason for the bad fit could be due to the presence of a degree of independent variability between the soft and hard bands.

\subsection{Frequency-dependent lag and coherence}
To probe the physical processes dominating on various timescales, we evaluate the time lag as a function of temporal frequency using the Fourier method described in \citet{vn97,no99}. The time lag between two different energy bands is given by the formula $\tau(\nu)=\phi(\nu)/2\pi \nu$, where $\phi(\nu)={\rm arg}(<C(\nu)>)$, is the phase of the average cross power spectrum, $C(\nu)$. The expression for the cross power spectrum is $C(\nu)=S^{*}(\nu)H(\nu)$, where $S(\nu)$ and $H(\nu)$ are the discrete Fourier transforms of two different time series $s(t)$ and $h(t)$, respectively. First, we composed light curves in three different energy bands: $E_{1}=0.3-0.8$\keV{} (super-soft band dominated by the free-free emission within the high-density disc), $E_{2}=0.8-1$\keV{} (soft band corresponding to the Fe-L line and dominated by the disc reflection) and $E_{3}=1.5-5$\keV{} (primary power-law continuum dominated). In order to obtain evenly-sampled light curves, we used the time bin size of $400$\s{}. We computed time lag by averaging the cross power spectra over five unfolded segments of time series and then averaging in logarithmically spaced frequency bins (each bin spans $\nu\rightarrow 1.5\nu$). The resulting lags between $E_{1}$ and $E_{2}$ and between $E_{1}$ and $E_{3}$ as a function of temporal frequency are shown in the left and right panels of Figure~\ref{lag_fre}, respectively. The lags were estimated relative to the $E_{1}$ and $E_{2}$ bands and the positive lag implies that the $E_{1}$ and $E_{2}$ bands are leading the $E_{3}$ variations. In the lowest frequency range $\nu\sim[2-6]\times10^{-5}$\hz{}, we detected a hard lag of $1173\pm327$\s{} and $815\pm267$\s{} between the 0.3$-$0.8\keV{} and 1.5$-$5\keV{} bands, and between the 0.8$-$1\keV{} and 1.5$-$5\keV{} bands, respectively. Therefore, at the lowest frequencies ($\leq 6\times10^{-5}$\hz{}) or longer timescales ($\geq16.7$\ks{}), the primary emission dominated hard band, lags behind the reflection dominated super-soft band (0.3$-$0.8\keV{}) by $1173\pm327$\s{} and soft band (0.8$-$1\keV{}) by $815\pm267$\s{}. However, as the frequency increases, the time lag between 0.3$-$0.8\keV{} and 1.5$-$5\keV{} becomes zero. At frequencies $\nu\sim[1-2]\times10^{-4}$\hz{}, the soft band (0.8$-$1\keV{}) dominated by the relativistic disc reflection, lags behind the primary power-law dominated hard band (1.5$-$5\keV{}), by $-183\pm145$\s{}. 

We examined the significance of the lags by calculating the Poisson noise subtracted coherence as a function of frequency, following the prescription of \citet{vn97}. The resulting coherence is high ($>0.5$) over the entire frequency range as shown in Figure~\ref{coh_fre}. The high coherence implies that time series in all three energy bands are well correlated linearly and the physical processes responsible for the soft X-ray excess, Fe emission complex and primary power-law emission are linked with each other.

\section{Energy-dependent variability}
\label{sec:EDV}
In order to study the energy dependence of variable spectral components, we determined the combined spectral and timing characteristics of Mrk~1044 with the use of different techniques: lag-energy spectrum, coherence spectrum and rms variability spectra on various timescales. 

\subsection{Lag-energy and coherence spectra}
The variation of the time lag as a function of energy can compare the energy spectral components and their relative lag and hence provides important insights into the origin of different spectral components. First, we extracted light curves in ten different energy bands: 0.3$-$0.4\keV{}, 0.4$-$0.5\keV{}, 0.5$-$0.6\keV{}, 0.6$-$0.8\keV{}, 0.8$-$1\keV{}, 1$-$1.5\keV{}, 1.5$-$2\keV{}, 2$-$3\keV{}, 3$-$5\keV{} and 5$-$10\keV{}. Then, we estimated the lag between the time series in each energy band and a reference band. The reference band is defined as the full (0.3$-$10\keV{}) band minus the energy band of interest. This allows us to obtain high signal-to-noise ratio in the reference band and also to overcome correlated Poisson noise. Here, the positive lag means that the variation in the chosen energy band is delayed relative to the defined reference band. In the left panel of Figure~\ref{lag_E}, we have shown the lag-energy spectrum of Mrk~1044 at the lowest frequency range $\nu\sim[2-6]\times10^{-5}$\hz{}, where the hard lag was observed. The low-frequency lag is increasing with energy and the lag-energy profile can be well described by a power-law of the form, $\tau(E)=1036.9\times E^{0.75}-912.2$\s{} and is shown as the solid, green line in Fig.~\ref{lag_E} (left). The low-frequency lag-energy spectrum of Mrk~1044 is similar to that observed in other AGN (e.g. Mrk~335 and Ark~564: \citealt{ka13}, MS~2254.9$-$3712: \citealt{al15}).

The lag-energy spectrum of Mrk~1044 in the high-frequency range $\nu\sim[1-2]\times10^{-4}$\hz{}, where we found a hint of the soft lag, is shown in the right panel of Fig.~\ref{lag_E}. The lag-energy profile has a peak at 0.8$-$1\keV{}, which could indicate a larger contribution of the reprocessed soft X-ray excess emission from the ionized accretion disc that results in a delayed emission with respect to the direct nuclear emission. This is consistent with our energy spectral fitting. However, we do not see any Fe-K lag because of poor signal-to-noise in the hard band. 

The noise-corrected coherence as a function of energy for these two frequency ranges, $\nu\sim[2-6]\times10^{-5}$\hz{} and $\nu\sim[1-2]\times10^{-4}$\hz{}, are shown in the left and right panels of Figure~\ref{coh_E}, respectively. The coherence is nearly consistent with unity over the entire energy range, indicating that the physical processes dominating different energy bands are well connected with each other.

\subsection{The absolute and fractional rms variability spectra}
The variation of the rms variability amplitude as a function of energy will allow us to distinguish between the constant and variable spectral components and is worthwhile to understand the origin of energy-dependent variability in accreting systems \citep{gi05,mi07,fa12,ma16,md17,ma17}. Moreover, the modelling of the fractional rms spectrum can quantify the percentage of variability of variable spectral components and probe the causal connection between them. 

\subsubsection{Deriving the absolute and fractional rms spectra}
We derived the \xmm{}/EPIC-pn (0.3$-$10\keV{}) frequency-averaged ($\sim 7.8\times10^{-6}-1.25\times10^{-3}$\hz{}) absolute and fractional rms variability spectra of Mrk~1044 following the prescription of \citet{va03}. The absolute rms ($\sigma_{\rm rms}$) is the square root of the excess variance ($\sigma_{\rm XS}^2$) and fractional rms ($F_{\rm var}$) is defined as the square root of the normalized excess variance $\sigma_{\rm NXS}^2$ which is the ratio of the excess variance ($\sigma_{\rm XS}^2$) and the square of the mean ($\overline{x}$) of the time series: 

\begin{equation}
\sigma_{\rm rms}=\sqrt{\sigma_{\rm XS}^2}=\sqrt{S^2-\overline{\sigma_{\rm err}^2}}
 \label{rms}
\end{equation}

\begin{equation}
 F_{\rm var}=\sqrt{\sigma_{\rm NXS}^2}=\sqrt{\frac{\sigma_{\rm XS}^{2}}{\overline{x}^2}}=\sqrt{\frac{S^2-\overline{\sigma_{\rm err}^2}}{\overline{x}^2}},
 \label{fvar}
\end{equation}
where $\sigma_{\rm XS}^2$ is defined by the sample variance $S^2$ minus the mean squared error $\overline{\sigma_{\rm err}^2}$.
The sample variance, $S^2$ is defined by
\begin{equation} 
S^{2} = \frac{1}{N-1}\sum\limits^{N}_{i=1}(x_{\rm i} - \overline{x})^{2}, 
\end{equation}
and mean squared error, $\overline{\sigma_{\rm err}^2}$ is:
\begin{equation}
\overline{\sigma_{\rm err}^{2}}=\frac{1}{N}\sum\limits_{i=1}^{N}\sigma_{\rm err,i}^{2}.
\end{equation}

The uncertainty on $F_{\rm var}$ was estimated using the equation~(B2) of \citet{va03} and is given by:
\begin{equation}
{\rm err}(F_{\rm var})=\frac{1}{2F_{\rm var}}\sqrt{\Big(\sqrt{\frac{2}{N}}.\frac{\overline{\sigma_{\rm err}^{2}}}{\overline{x}^{2}}\Big)^{2}+\Big(\sqrt{\frac{\overline{\sigma_{\rm err}^{2}}}{N}}.\frac{2F_{\rm var}}{\overline{x}}\Big)^{2}}
\label{fvar_err}
\end{equation}

We derived the deadtime corrected, background subtracted EPIC-pn light curves in 19 different energy bands with the time bins of 400\s{}. Then we computed the frequency-averaged ($\sim 7.8\times10^{-6}-1.25\times10^{-3}$\hz{}) absolute rms ($\sigma_{\rm rms}$) in units of counts~s$^{-1}$~keV$^{-1}$ and fractional rms ($F_{\rm{var}}$) in each time series. We also derived the EPIC-pn mean spectrum by rebinning the response matrix with same energy binning as the rms spectra. We have shown the EPIC-pn mean and absolute rms spectra in the left panel and fractional rms spectrum in the right panel of Figure~\ref{rms_data}. The fractional rms amplitude of the source decreases with energy convexly with a hint of drops at $\sim0.7$\keV{} and $\sim0.9$\keV{}. 

Then we explored the frequency-resolved variability spectra of Mrk~1044 by calculating the absolute and fractional rms variability amplitudes in two broad frequency bands: $\nu_{\rm{low}}\sim[1.7-10]\times10^{-5}$\hz{} and $\nu_{\rm{high}}\sim[1-10]\times10^{-4}$\hz{}. The corresponding timescales are $\sim 5-60$\ks{} and $\sim 0.5-10$\ks{}, respectively. To obtain the low-frequency variability spectra, we extracted the corrected light curves with the time bin size of $\Delta t=5$\ks{} and segment length of $t=60$\ks{}. For the high-frequency variability spectra, the chosen time bin size and segment length of the time series were $\Delta t=500$\s{} and  $t=10$\ks{}, respectively. The derived low and high-frequency absolute and fractional rms spectra of Mrk~1044 are shown in the left and right panels of Figure~\ref{rms_nu_resol} and Figure~\ref{fvar_nu_resol}, respectively. 

\subsubsection{Modelling the fractional rms spectrum}
To identify the variable spectral components responsible for the observed energy-dependent variability of Mrk~1044, we modelled the 0.3$-$10\keV{} $F_{\rm var}$ spectrum following the approach of \citet{ma17}. The $F_{\rm var}$ spectrum can be considered as a representative of the fractional rms variations of the average energy spectrum. The best-fitting average energy spectral model of Mrk~1044 consists of a direct power-law continuum ($f_{\rm DPC}$), an ionized high-density relativistic reflection continuum ($f_{\rm RRC}$) and a neutral distant reflection ($f_{\rm dist}$), modified by the Galactic ($f_{\rm GA}$) and intrinsic ($f_{\rm abs}$) absorption components and can be expressed as

\begin{equation}
f(E)=f_{\rm GA}f_{\rm abs}[f_{\rm DPC}(E)+f_{\rm RRC}(E)+f_{\rm dist}(E)]
\end{equation}

If we consider that the observed energy-dependent variability of Mrk~1044 is caused by variations in the normalization ($N_{\rm PL}$) and photon index ($\Gamma$) of the hot Comptonization component as well as in the normalization ($N_{\rm blur}$) of the relativistic reflection component, then the variations in the average energy spectrum can be written as

\begin{small}
\begin{equation}
\Delta f(E)=\Delta f_{\rm DPC}(E,N_{\rm PL},\Gamma)+\Delta f_{\rm RRC}(E,N_{\rm blur})
\end{equation}
\end{small}
where
\begin{small}
\begin{equation}
\Delta f_{\rm DPC}(E,N_{\rm PL},\Gamma)=f_{\rm DPC}(N_{\rm PL}+\Delta N_{\rm PL},\Gamma+\Delta \Gamma)-f_{\rm DPC}(N_{\rm PL},\Gamma)
\label{eq1}
\end{equation}
\end{small}
and
\begin{small}
\begin{equation}
\Delta f_{\rm RRC}(E,N_{\rm blur})=f_{\rm RRC}(N_{\rm blur}+\Delta N_{\rm blur})-f_{\rm RRC}(N_{\rm blur})
\label{eq2}
\end{equation}
\end{small}

Therefore, we can write the expression for the fractional rms spectrum $F_{\rm var}(E)$ using the equation~(3) of \citet{ma17}:
\begin{small}
\begin{equation}
 F_{\rm var}(E)=\frac{\sqrt{<(\Delta f_{\rm DPC}(E,N_{\rm PL},\Gamma)+\Delta f_{\rm RRC}(E,N_{\rm blur}))^2>}}{f_{\rm DPC}(E)+f_{\rm RRC}(E)+f_{\rm dist}(E)}.
\label{eq3}
\end{equation}
\end{small}
We simplified the expression for the fractional rms spectrum by expanding the first term on the right-hand side of equation~(\ref{eq1}) and (\ref{eq2}) in a Taylor series around the variable parameters ($N_{\rm PL}$, $\Gamma$), $N_{\rm blur}$, respectively and neglecting the higher order (second-order derivatives onward) terms. We can also obtain the correlation coefficients, $\alpha$ between $\Delta N_{\rm PL}$ and $\Delta \Gamma$ and $\beta$ between $\Delta N_{\rm PL}$ and $\Delta N_{\rm blur}$ from the numerator of equation~(\ref{eq3}). We then fitted the observed fractional rms spectrum of Mrk~1044 by implementing the equation~(\ref{eq3}) in \textsc{ISIS}~v.1.6.2-40 \citep{ho00} as a local model.

Initially, we fitted the 0.3$-$10\keV{} $F_{\rm var}$ spectrum of Mrk~1044 with a model consisting of variable hot Comptonization component ($\Delta f_{\rm DPC}$) with free parameters $\Delta N_{\rm PL}$, $\Delta \Gamma$. We also consider that $\Delta N_{\rm PL}$ and $\Delta \Gamma$ are correlated by the correlation coefficient $\alpha$. This provided a statistically unacceptable fit with $\chi^{2}$/d.o.f = 80.1/16. Then we included variability in the normalization ($\Delta N_{\rm blur}$) of the ionized disc reflected emission and also consider that variations in the normalization of the direct power-law continuum, $\Delta N_{\rm PL}$ and inner disc reflection, $\Delta N_{\rm blur}$ are connected to each other by the correlation coefficient, $\beta$. This model describes the observed fractional rms spectrum of Mrk~1044 well with $\chi^{2}$/d.o.f = 12.8/14 ($\Delta\chi^{2}$=$-67.3$ for 2 d.o.f). The best-fitting values for the fractional rms spectral model parameters are: $\frac{\Delta N_{\rm PL}}{N_{\rm PL}}=62.8^{+12.6}_{-11.6}$~per~cent, $\frac{\Delta \Gamma}{\Gamma}=8.6^{+5.4}_{-2.5}$~per~cent, $\alpha=1.0^{+0p}_{-0.21}$, $\frac{\Delta N_{\rm blur}}{N_{\rm blur}}=25.0^{+2.5}_{-3.0}$~per~cent and $\beta=-0.34^{+0.34}_{-0.33}$. The 0.3$-$10\keV{} frequency-averaged fractional rms variability spectrum, the best-fitting variability spectral model and the data-to-model ratio are shown in Figure~\ref{rms_model}. Thus we infer the presence of a less variable inner disc reflection with variable flux and a more variable direct coronal emission where the flux variations of the coronal emission and relativistic disc reflection are either uncorrelated or moderately anti-correlated with each other.  

\section{Summary and Discussion}
\label{sec:discussion}
We present the first results from the broadband (0.3$-$50\keV{}) spectral and timing studies of the highly accreting, narrow-line Seyfert~1 galaxy Mrk~1044 using \xmm{} ($\sim130$\ks{}), \nustar{} ($\sim22$\ks{}) and \swift{} ($\sim29$\ks{}) observations. Here we perform time-averaged spectral modelling, frequency and energy-dependent time-lag, coherence, absolute and fractional rms variability spectral analyses. We investigate the underlying physical processes (Comptonization, reverberation and propagation fluctuation) surrounding the SMBH and disentangle various emission components responsible for the observed energy-dependent variability on various timescales. The main results of our work are summarized below: 

\begin{enumerate}
\item The time-averaged \xmm{} (0.3$-$10\keV{}) spectrum of Mrk~1044 shows strong soft X-ray excess emission below $\sim2$\keV{}, a narrow Fe~K$_{\alpha}$ emission line at $\sim6.4$\keV{}, a broad Fe emission line at $\sim6.84$\keV{} and a possible O~VIII wind with an outflowing velocity of $(0.1\pm0.01)c$. 

\item The broadband (0.3$-$50\keV{}) quasi-simultaneous \swift{} and \nustar{} spectra confirm the presence of a soft X-ray excess below $\sim1$\keV{}, a narrow Fe~K$_{\alpha}$ emission line at $\sim6.4$\keV{} and reveal a Compton hump at $\sim15-30$\keV{}.

\item The fitting of the average energy spectrum requires an ionized high-density relativistic reflection model with a broken power-law emissivity profile of the accretion disc, to describe the soft X-ray excess, broad Fe line and Compton hump. The best-fitting values for the electron density, inner radius and break radius of the disc are $n_{\rm e}=5.2^{+5.2}_{-4.2}\times10^{16}$~cm$^{\rm -3}$, $r_{\rm in}=1.31^{+0.08}_{-0.05}r_{\rm g}$ and $r_{\rm br}=3.2^{+0.1}_{-0.1}r_{\rm g}$, respectively.

\item During the 2013 \xmm{} observation, Mrk~1044 shows a significant variability with changes in the total X-ray (0.3$-$10\keV{}) count rate by $\sim50$~per~cent on a timescale of $\sim800$\s{}. The hardness$-$flux analysis shows that source hardness decreases with the total X-ray (0.3$-$10\keV{}) count rate, suggesting a softer-when-brighter trend as usually observed in radio-quiet narrow-line Seyfert~1 AGN (Fig.~\ref{hr_flux}).

\item At low frequencies ($\sim[2-6]\times10^{-5}$\hz{}), the hard band (1.5$-$5\keV{}) which is dominated by the illuminating continuum lags behind the relativistic reflection dominated super-soft (0.3$-$0.8\keV{}) and soft (0.8$-1$\keV{}) bands by ($1173\pm327$)\s{} and ($815\pm267$)\s{}, respectively. As the frequency increases, the time-lag between the super-soft and hard bands disappears. However, we do see a negative lag between the soft (0.8$-$1\keV{}) and hard (1.5$-$5\keV{}) bands where the soft band lags behind the hard band by ($183\pm145$)\s{} at higher frequencies ($\sim[1-2]\times10^{-4}$\hz{}) (Fig.~\ref{lag_fre}).

\item The low-frequency lag-energy spectrum is featureless and has a power-law like shape, while the high-frequency lag spectrum has a maximum value at 0.8$-$1\keV{} where the Fe-L emission line peaks and can be attributed to the delayed emission from the inner accretion disc (Fig.~\ref{lag_E}). However, we do not see any Fe-K lag. The non-detection of the Fe-K lag has previously been reported in another Seyfert~1 galaxy MCG--6-30-15 \citep{ka14}.

\item The source variability decreases with energy during both 2013 and 2016 observations. The fractional rms amplitudes in the 0.3$-$2\keV{}, 2$-$10\keV{}, 3$-$10\keV{} and 10$-$50\keV{} bands are $F_{\rm var,0.3-2}=(20.3\pm0.1$)~per~cent, $F_{\rm var,2-10}=(15.5\pm0.5$)~per~cent, $F_{\rm var,3-10}=(15.1\pm1.5$)~per~cent, $F_{\rm var,10-50}=(7.8\pm4.8$)~per~cent, respectively. The modelling of the \xmm{}/EPIC-pn frequency-averaged ($\sim 7.8\times10^{-6}-1.25\times10^{-3}$\hz{}) fractional rms spectrum reveals that the observed energy-dependent variability of Mrk~1044 is mainly driven by two components: more variable illuminating continuum and less variable relativistic reflection from an ionized high-density accretion disc. 

\end{enumerate}

\begin{figure}
\includegraphics[width=0.47\textwidth,angle=-0]{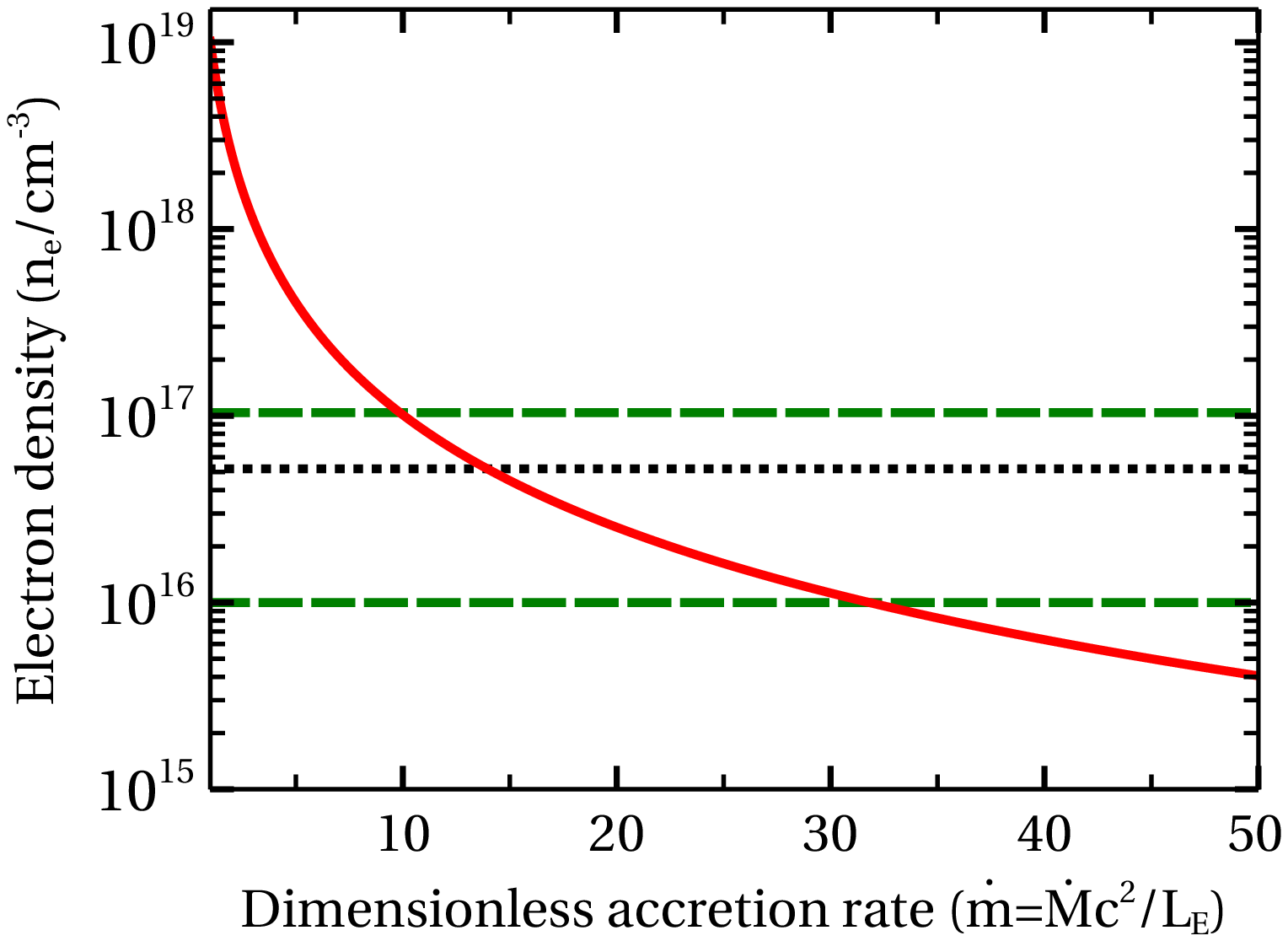}
\caption{The variation of the disc electron density as a function of the dimensionless mass accretion rate derived from the radiation pressure solution of \citet{sz94}, and considering $\alpha=0.1$, $M_{\rm BH}=3\times10^{6}M_{\odot}$, $f=0.9$ and $r=10$. The dotted and dashed lines represent the observed value of the electron density and its 90~per~cent confidence limits, respectively.}
\label{density_mdot}
\end{figure}

\subsection{Origin of the soft and hard X-ray excess emission: high-density disc reflection}
We investigate the origin of the soft and hard X-ray excess emission including the Fe~K emission complex and Compton hump in this low-mass, highly accreting AGN. The modelling of the hard band (3$-$10\keV{}) EPIC-pn spectrum of the source requires a hot Comptonization component with a photon index of $\Gamma=2.25^{+0.07}_{-0.09}$ along with an ionized, relativistic reflection with a single power-law emissivity profile for the accretion disc to fit the broad Fe emission line and a neutral, non-relativistic reflection from the distant medium to fit the narrow Fe~K$_{\alpha}$ emission line. The extrapolation of the 3$-$10\keV{} best-fitting model down to 0.3\keV{} reveals strong residuals below $\sim2$\keV{} due to the soft X-ray excess emission. The modelling of the full band (0.3$-$10\keV{}) EPIC-pn spectrum including the soft X-ray excess requires a broken power-law emissivity profile and high electron density ($n_{\rm e}\sim5\times10^{16}$~cm$^{\rm -3}$) for the accretion disc. Moreover, the photon index ($\Gamma\sim2.25$), disc inclination angle ($\theta\sim45^{\circ}$), inner emissivity index ($q_{\rm in}\sim8-10$) obtained from considering only the 3$-$10\keV{} spectrum explaining the Fe emission complex, are consistent with that inferred from fitting the full band (0.3$-$10\keV{}) spectrum including soft X-ray excess. We find that the shape of the inner emissivity profile ($\epsilon_{\rm in}\propto r^{-9}$ for $r<r_{\rm br}$) describing the broad Fe emission line remains unaffected even after the inclusion of the soft band data and we only require another flat power-law emissivity profile ($\epsilon_{\rm out}\propto r^{-2}$ for $r>r_{\rm br}$) of the disc to explain the soft X-ray excess. The physical implication of the broken power-law emissivity profile is that the disc emitting regions responsible for the broad Fe emission line and soft X-ray excess are different and separated by the break radius. The best-fit value of the break radius obtained from the full band (0.3$-$10\keV{}) EPIC-pn spectrum is $r_{\rm br}\sim3r_{\rm g}$ which implies that soft X-ray excess originates from the high-density accretion disc above $\sim3r_{\rm g}$ and the broad Fe emission line arises in the innermost regions of the accretion disc below the break radius of around $3r_{\rm g}$. 

We also detect a Compton hump at around 15$-$30\keV{} during the 2016 \nustar{} observation. The inclusion of the 10$-$50\keV{} spectral data does not affect the spectral model parameters (disc inclination angle, spin, emissivity indices, break radius) obtained from fitting of the 0.3$-$10\keV{} spectrum only. The best-fit value of the disc electron density as derived from the modelling of the broadband (0.3$-$50\keV{}) spectral data is $n_{\rm e}=5.2^{+5.2}_{-4.2}\times10^{16}$~cm$^{\rm -3}$. Mrk~1044 is known to be a highly accreting AGN with the dimensionless mass accretion rate of $\dot{m}=\frac{\dot{M}c^{2}}{L_{\rm E}}=16.6^{+25.1}_{-10.1}$ \citep{du15}. At high accretion rate, the inner region of a standard $\alpha$-disc \citep{ss73} is radiation pressure-dominated and the electron density of the disc can be written as \citep{sz94}

\begin{equation}
n_e=\frac{1}{\sigma_{\rm T}R_{\rm S}} \frac{256\sqrt{2}}{27}\alpha^{-1}r^{3/2}\dot m^{-2} [1-(3/r)]^{-1} (1-f)^{-3}.
\label{density}
\end{equation}
where $\sigma_{\rm T}=6.64\times10^{-25}$~cm$^2$ is the Thomson scattering cross section, $R_{\rm S}=2GM_{\rm BH}/c^{2}$ is the Schwarzschild radius, $M_{\rm BH}$ is the black hole mass, $\alpha=0.1$ is the disc viscosity parameter, $r=R/R_{\rm S}$, $R$ is the characteristic disc radius, $\dot{m}=\frac{\dot{M}c^{2}}{L_{\rm E}}$ is the dimensionless mass accretion rate, $f$ is the fraction of the total power released by the disc into the corona. The variation of the disc electron density ($n_e$) with the dimensionless mass accretion rate ($\dot{m}$) for $\alpha=0.1$, $M_{\rm BH}=3\times10^{6}M_{\odot}$, $f=0.9$ and $r=10$ is shown as the solid curve in Figure~\ref{density_mdot}. As evident from Fig.~\ref{density_mdot}, the assumption of constant disc density ($n_{\rm e}=10^{15}$~cm$^{\rm -3}$) is not physically realistic for low-mass AGN even when the mass accretion rate is very high. The observed best-fit value for the disc electron density of the source and its 90~per~cent confidence limits are shown as the dotted and dashed lines in Fig.~\ref{density_mdot}, respectively. The corresponding dimensionless mass accretion rate of Mrk~1044 estimated using equation~(\ref{density}) is $\dot{m}\approx10-32$, which is in agreement with that found by \citet{du15}. We further verified the SMBH mass with the use of X-ray variability techniques as pioneered by \citet{po12}. The relation between the SMBH mass ($M_{\rm BH,7}$) in units of $10^{7}M_{\odot}$ and normalized excess variance ($\sigma_{\rm NXS}^{2}$) in the 2$-$10\keV{} light curves of 10\ks{} segments and the bin size of 250\s{}, can be written as
\begin{equation}
\log(\sigma_{\rm NXS}^{2})=(-1.83\pm0.1)+(-1.04\pm0.09)\log(M_{\rm BH,7}).
\label{mass_Fvar}
\end{equation}
The SMBH mass of Mrk~1044 measured using equation~(\ref{mass_Fvar}) is $M_{\rm BH}=(4-5)\times10^{6}M_{\odot}$ which is close to that measured by \citet{du15}.

We detected a broad absorption line with the rest frame energy of around $0.72$\keV{} as a single feature and one of the strongest lines present both in the EPIC-pn and RGS spectra. The modelling of the RGS spectrum with the photoionized absorption model infers the presence of an O~VIII wind moving with the outflow velocity of $(0.1\pm0.01)c$. The existence of a persistent (flux independent) O~VIII outflow has previously been reported in another highly-accreting NLS1 AGN IRAS~13224--3809 \citep{pin18}. 

Our broadband spectral analysis also reveals that the central SMBH of Mrk~1044 is spinning very fast with the spin parameter $a=0.997^{+0.001}_{-0.016}$. The high spin ($a\approx1$) is a natural consequence of high radiative efficiency of prograde accretion (see Fig.5 of \citealt{ki08}), which is further supported by the very high accretion rate of Mrk~1044. So the high spin is not the result of the shortcoming of the relativistic reflection model. Hence we conclude that the origin of the soft X-ray excess, broad Fe emission line and Compton hump is the relativistic reflection from an ionized, high-density accretion disc around the rapidly rotating SMBH.

\subsection{Probing Comptonization, propagation fluctuation and reverberation scenarios}
We detected hard lags where the $1.5-5$\keV{} band emission lags behind the emission from the 0.3$-$0.8\keV{} and 0.8$-$1\keV{} bands by $1173\pm327$\s{} and $815\pm267$\s{}, respectively in the lowest frequency range $(2-6)\times10^{-5}$\hz{}. Here we explore both the Compton scattering and propagation fluctuation scenarios for the origin of the hard lag. In the framework of Compton up-scattering, a soft X-ray photon with energy $E_{\rm S}$ after $N$ scatterings produces a hard X-ray photon of energy $E_{\rm H}$ \citep{zd85}
\begin{equation}
E_{\rm H}=A^{N}E_{\rm S}, 
\label{comp}
\end{equation}
where $A=1+4\theta+16\theta^{2}$, $\theta=\frac{k_{B}T_{e}}{m_{e}c^{2}}$ and $T_{e}$ and $m_{e}$ are the electron temperature and rest mass, respectively. 
If $t_{c}$ is the time delay between successive scatterings and the hard X-ray photons are produced within the optically thin, hot corona after $N$ scatterings, then the Comptonization timescale is
\begin{equation}
t_{comp}=Nt_{c}.
\end{equation}
For an optically thin, hot ($T_{e}\sim100$\keV{}) corona of size $r\sim10r_g$, the Comptonization timescales of $\sim0.3$\keV{} and $\sim0.8$\keV{} photons which eventually produce $\sim5$\keV{} photons within the corona, are $\sim470$\s{} and $\sim300$\s{}, respectively. The estimated Comptonization timescales are much smaller than the observed hard lags. Therefore, the origin of the hard lag is not entirely due to the Compton up-scattering of the soft X-ray photons from the disc. Then we investigated the viscous propagation fluctuation scenario as proposed by \citet{ko01}. In this scenario, the fluctuations produced at different radii in the disc propagates inwards and then hits the edge of the X-ray emitting corona on timescales corresponding to the viscous timescale. The viscous timescale at the disc emission radius $r$ is defined by
\begin{equation}
t_{\rm vis} = \frac{1}{\alpha}\left(\frac{r}{h}\right)^{2}t_{\rm dyn},
\end{equation}
where $\alpha\sim0.1$ is the viscosity parameter, $h$ is the disc height which is much smaller than the disc radius $r$ for a standard thin disc. However, in the inner regions of the disc, $r$ is only a few $r_{\rm g}$ and one can consider $h\approx r$. Therefore the expression for the viscous timescale becomes $t_{\rm vis}\approx\frac{1}{\alpha}t_{\rm dyn}$. For a SMBH of mass $M_{\rm BH}$, the dynamical timescale is $t_{\rm dyn}=\large\sqrt{\frac{r^3}{GM_{\rm BH}}}\approx500\left(\frac{M_{\rm BH}}{10^{8}M_{\rm \odot}}\right)\left(\frac{r}{r_{\rm g}}\right)^{3/2}$\s{}. From the modelling of the full band \xmm{}/EPIC-pn spectrum with the relativistic reflection model, we found that the soft X-ray photons are produced in the disc above the break radius of $r_{\rm br}=r\sim3r_{\rm g}$. Therefore, the minimum viscous timescale required for a soft X-ray photon to hit the edge of the corona is estimated to be $t_{\rm vis}\sim780$\s{}. Thus the observed hard lags of $1173\pm327$\s{} and $815\pm267$\s{} between the super-soft ($0.3-0.8$\keV{}) and hard ($1.5-5$\keV{}) bands and between the soft ($0.8-1$\keV{}) and hard ($1.5-5$\keV{}) bands can indeed correspond to the sum of the  fluctuation propagation timescale ($\sim780$\s{}) between the inner disc and edge of the corona and the Comptonization timescales ($\sim470$\s{} and $\sim300$\s{}) of the $\sim0.3$\keV{} and $\sim0.8$\keV{} seed photons within the corona, respectively. In the propagation fluctuation scenario, we expect to see log-linear behaviour of the lag-energy spectrum. However, the shape of the low-frequency lag-energy spectrum is not entirely consistent with the log-linear lags. The possible reason for the dilution of the log-linear behaviour could be the Compton up-scattering of the soft X-ray photons in the corona \citep{ut11}. Thus we conclude that both propagation fluctuation and Comptonization are responsible for the observed time-lags on longer timescales.

We also detected a soft lag where the 0.8$-$1\keV{} band emission lags behind the harder 1.5$-$5\keV{} band emission by $38-328$\s{} in the higher frequency range $\sim(1-2)\times10^{-4}$\hz{}. We estimated the soft lag for Mrk~1044 using the scaling relation between the soft lag amplitude ($|\tau|$) and SMBH mass ($M_{\rm BH}$) from \citet{dm13}: $\log|\tau| = 1.98[\pm0.08] + 0.59[\pm0.11]\log(M_{\rm BH})$, where $M_{\rm BH}$ is the SMBH mass in units of $10^{7}M_{\odot}$. For Mrk~1044, the scaling relation suggests the amplitude of the soft lag to lie in the range $|\tau|=44-49$\s{}, which is consistent with our measured time-lag. The origin of the soft lag can be explained in the context of the reverberation scenario where the soft X-ray excess emission is produced due to relativistic reflection from an ionized accretion disc and lag amplitude corresponds to the light-crossing time between the direct power-law emitting hot corona and soft X-ray emitting inner disc. For Mrk~1044, the measured soft lag of ($38-328$)\s{} implies that the physical separation between the corona and soft X-ray emitting inner disc is in the range $r=(2.5-22)r_{\rm g}$. This radius is consistent with our average energy spectral fitting with the high-density relativistic reflection model, which revealed that the soft X-ray excess emission was originated above a certain radius ($r_{\rm br}\sim3r_{\rm g}$), known as the break radius of the accretion disc. However, we do not find the Fe-K lag in Mrk~1044 which could be caused by several reasons. Since Mrk~1044 is a low-mass AGN and Fe-K emission originates from the innermost region ($r\sim1.3r_{\rm g}$) of the disc, the predicted Fe-K lag amplitude is only $\sim20$\s{} which is difficult to detect given the signal-to-noise in the hard band. Another reason for the non-detection of the Fe-K lag could be due to the absence of correlated variability between the direct continuum and disc reflection or changes in the height or radius of the corona, which we discuss in the next section.

\subsection{Origin of the broadband X-ray variability and disc/corona connection}
The X-ray emission from the source is variable and the variability amplitude is both energy and frequency-dependent. The source showed a decrease in fractional rms amplitude with energy and the shape of the variability spectrum is very similar in both low ($\nu_{\rm low}\sim[1.7-10]\times10^{-5}$\hz{}) and high ($\nu_{\rm high}\sim[1-10]\times10^{-4}$\hz{}) frequency bands, although the amplitude is different on different timescales (see Fig.~\ref{fvar_nu_resol}). The fractional variability amplitudes of the source in the $0.3-10$\keV{}, $0.3-2$\keV{} and $2-10$\keV{} bands are $(15.9\pm0.1$)~per~cent, $(16.5\pm0.1$)~per~cent and $(11.1\pm0.4$)~per~cent, respectively on timescales of $\sim 5-60$\ks{} and $(17.1\pm0.1$)~per~cent, $(17.6\pm0.2$)~per~cent and $(13.4\pm0.5$)~per~cent, respectively on timescales of $\sim0.5-10$\ks{}. Thus, it is apparent that the X-ray emission from Mrk~1044 is more variable on short timescales ($<10$\ks{}) in all three energy bands, implying that the shorter timescale variability has been originated in a more compact emitting region which radiates at the smaller radii of the disc and hence close to the SMBH. The modelling of the frequency-averaged fractional rms spectrum reveals the presence of two-component variability: direct coronal and reflected inner disc emission, both of which are variable either in an uncorrelated or moderately anti-correlated manner and the disc reflection variability is about a factor of $2$ smaller than the coronal variability. The primary power-law emission from the hot corona is variable both in flux and spectral index. We find that the flux and spectral index variability are positively correlated with each other by $\sim79-100$~per~cent, which can be explained in the context of Compton cooling of the soft seed photons in the hot corona. The lack of positive correlation between the hot coronal and inner disc reflected emission indicates that either the primary X-ray emitting hot corona is compact and moving closer or further from the central SMBH or the corona is extended and is expanding or contracting with time (see \citealt{wi14}). In this scenario, the total number of photons emitted from the source remains constant and the observed energy-dependent X-ray variability is mainly caused by changes in the location or geometry of the corona in terms of height if the corona is compact or radius if the corona is extended.

\section{Acknowledgments}
We thank the anonymous referee for useful comments that improved the quality of the paper. LM gratefully acknowledges financial support from the University Grants Commission (UGC), Government of India. LM is highly grateful to the University of Cambridge for supporting an academic visit. WNA and ACF acknowledge support from the European Union Seventh Framework Programme (FP7/2013-2017) under grant agreement no. 312789, StrongGravity. WNA, ACF and CP acknowledge support from the European Research Council through Advanced Grant 340442, on Feedback. This research has made use of archival data of \xmm{}, \swift{} and \nustar{} observatories through the High Energy Astrophysics Science Archive Research Center Online Service, provided by the NASA Goddard Space Flight Center. This research has made use of the NASA/IPAC Extragalactic Database (NED), which is operated by the Jet Propulsion Laboratory, California Institute of Technology, under contract with the NASA. This research has made use of the XRT Data Analysis Software (XRTDAS) developed under the responsibility of the ASI Science Data Center (ASDC), Italy. This research has made use of the NuSTAR Data Analysis Software (NuSTARDAS) jointly developed by the ASI Science Data Center (ASDC, Italy) and the California Institute of Technology (Caltech, USA). This research has made use of ISIS functions (ISISscripts) provided by ECAP/Remeis observatory and MIT (http://www.sternwarte.uni-erlangen.de/isis/). Figures in this paper were made with the graphics package \textsc{pgplot} and GUI scientific plotting package \textsc{veusz}. This research made use of the \textsc{python} packages \textsc{numpy}, \textsc{scipy} and \textsc{matplotlib}.

\appendix
\section{Additional plots}
\label{sec:appendix}

\begin{figure*}
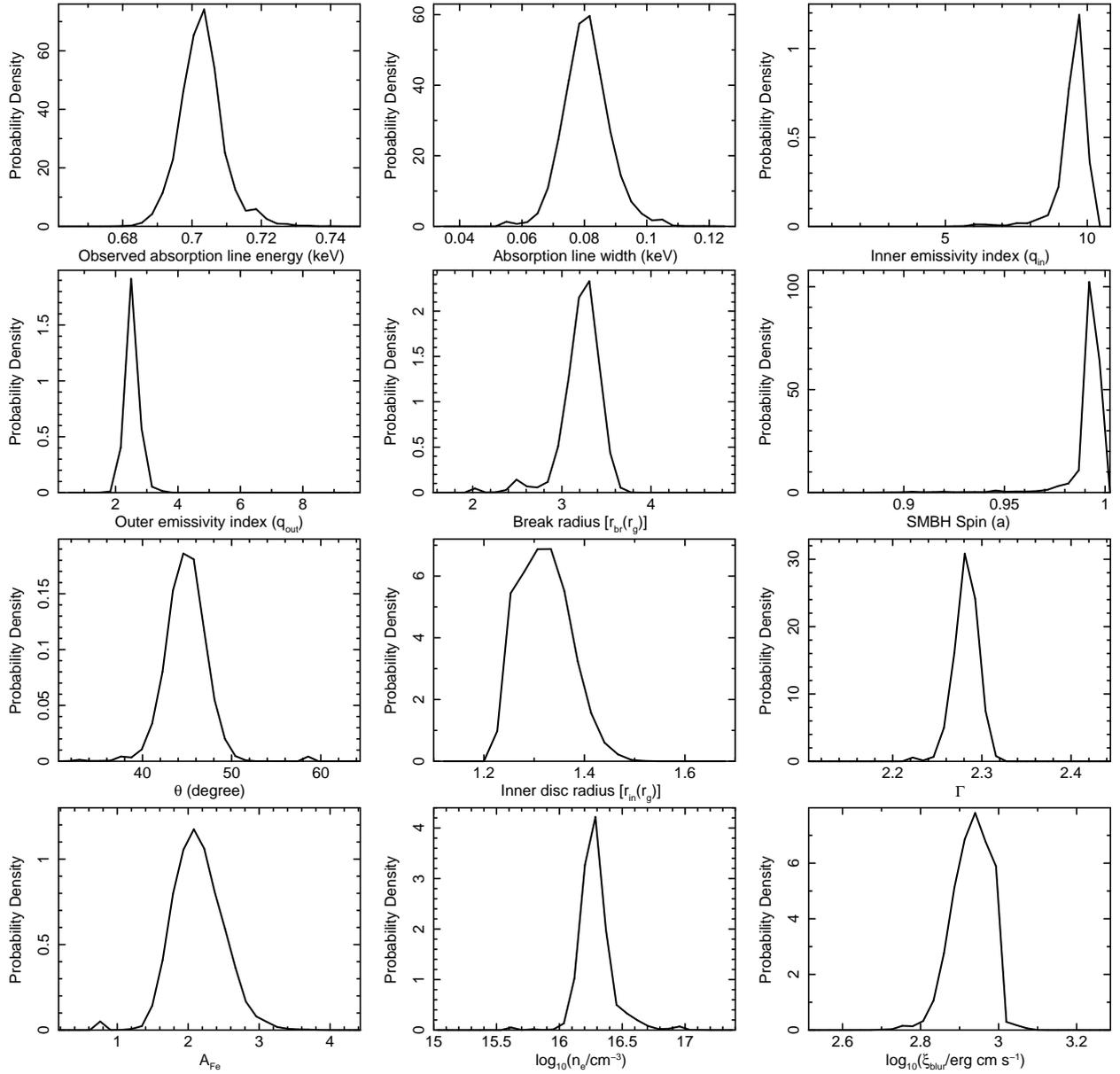

\includegraphics[scale=0.21,angle=-90]{fig4b.ps}
\includegraphics[scale=0.21,angle=-90]{fig4c.ps}
\includegraphics[scale=0.21,angle=-90]{fig4d.ps}
\includegraphics[scale=0.21,angle=-90]{fig4e.ps}
\includegraphics[scale=0.21,angle=-90]{fig4f.ps}
\includegraphics[scale=0.21,angle=-90]{fig4g.ps}
\includegraphics[scale=0.21,angle=-90]{fig4h.ps}
\includegraphics[scale=0.21,angle=-90]{fig4i.ps}
\includegraphics[scale=0.21,angle=-90]{fig4j.ps}
\includegraphics[scale=0.21,angle=-90]{fig4a.ps}
\includegraphics[scale=0.21,angle=-90]{fig4k.ps}
\includegraphics[scale=0.21,angle=-90]{fig4l.ps}
\caption{The probability distributions of the observed absorption line energy ($E_{\rm abs}$) and width ($\sigma_{\rm abs}$), inner ($q_{\rm in}$) and outer ($q_{\rm out}$) emissivity indices, break radius ($r_{\rm br}$), SMBH spin ($a$), disc inclination angle ($\theta^{\circ}$), inner disc radius ($r_{\rm in}$), photon index ($\Gamma$), Fe abundance ($A_{\rm Fe}$), disc electron density ($\log n_{\rm e}$) and ionization parameter ( $\log\xi_{\rm blur}$). The results are obtained from a $10^6$ length MCMC run on the best-fitting full band ($0.3-10$\keV{}) EPIC-pn spectral model parameters.}
\label{prob}
\end{figure*}

\begin{figure*}
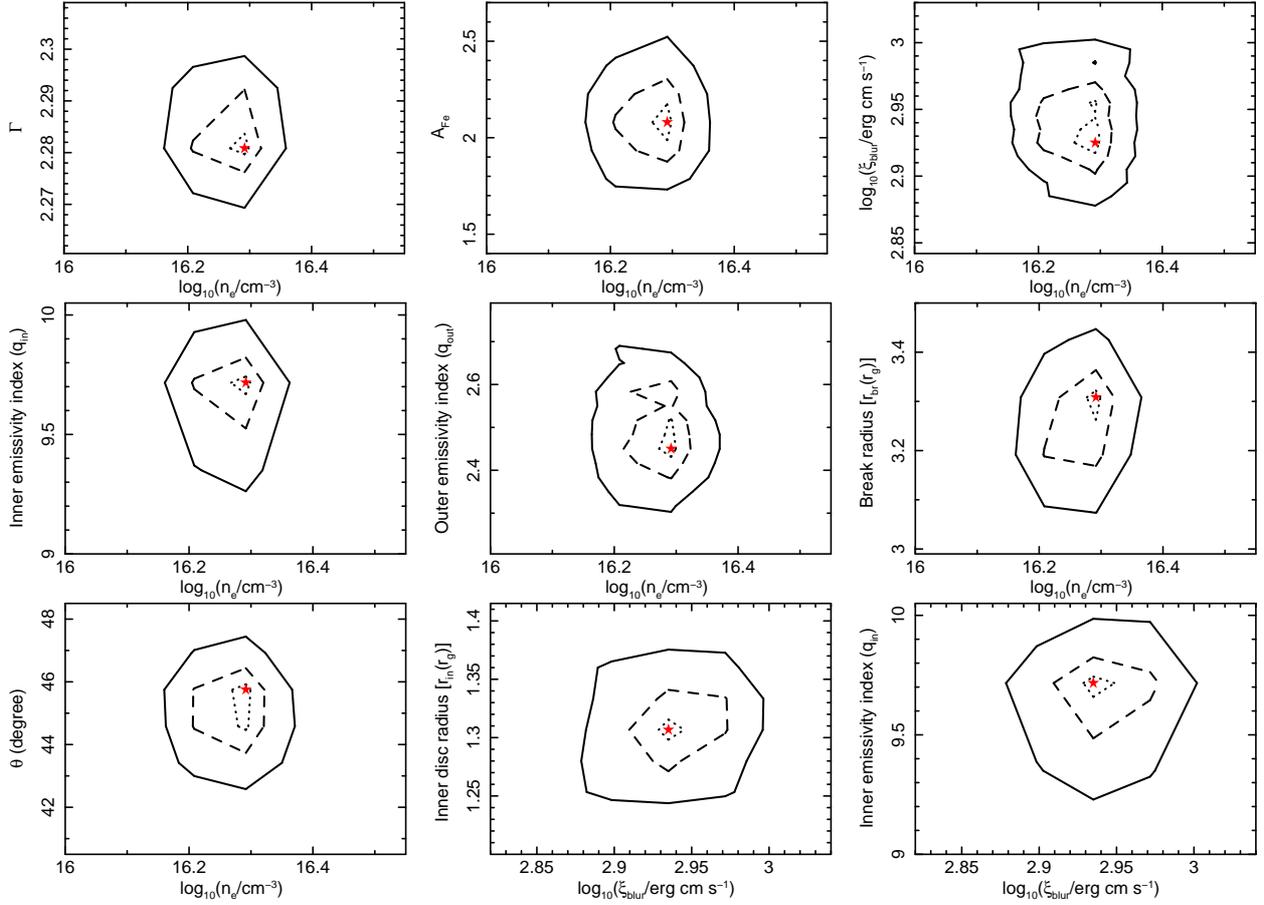

\includegraphics[scale=0.21,angle=-90]{fig5a.ps}
\includegraphics[scale=0.21,angle=-90]{fig5b.ps}
\includegraphics[scale=0.21,angle=-90]{fig5c.ps}
\includegraphics[scale=0.21,angle=-90]{fig5d.ps}
\includegraphics[scale=0.21,angle=-90]{fig5e.ps}
\includegraphics[scale=0.21,angle=-90]{fig5f.ps}
\includegraphics[scale=0.21,angle=-90]{fig5g.ps}
\includegraphics[scale=0.21,angle=-90]{fig5h.ps}
\includegraphics[scale=0.21,angle=-90]{fig5i.ps}
\caption{The contour plots between various spectral parameters for the full band ($0.3-10$\keV{}) EPIC-pn spectrum. The  dotted, dashed and solid lines correspond to 68, 90 and 99~per~cent confidence levels, respectively. The star symbols represent the best-fitting values of the parameters. All the parameters are well constrained and non-degenerate.}
\label{cont}
\end{figure*}

\bsp	
\label{lastpage}
\end{document}